\documentclass[twocolumn, journal]{IEEEtran}
\IEEEoverridecommandlockouts

\usepackage{bm}
\usepackage{color}
\usepackage{graphicx}
\usepackage{amsmath}
\allowdisplaybreaks[4]
\usepackage{amssymb}
\usepackage{algorithm}
\usepackage{algorithmic}
\usepackage{ulem}
\usepackage{lineno}
\usepackage{lettrine}
\usepackage{subfigure}
\usepackage{epstopdf}
\usepackage{stfloats}
\usepackage{multirow}
\usepackage{booktabs}
\usepackage{array}
\usepackage{amsthm}
\usepackage{caption}
\usepackage{subfigure}
\usepackage{setspace}
\usepackage{makecell}

\newcommand{\be}{\begin{equation}}
	\newcommand{\ee}{\end{equation}}
\newcommand{\bea}{\begin{eqnarray}}
	\newcommand{\eea}{\end{eqnarray}}
\newcommand{\ba}{\begin{array}}
	\newcommand{\ea}{\end{array}}

\def\BibTeX{{\rm B\kern-.05em{\sc i\kern-.025em b}\kern-.08em
    T\kern-.1667em\lower.7ex\hbox{E}\kern-.125emX}}
\usepackage{balance}

\makeatletter

\newcommand{\Rmnum}[1]{\expandafter\@slowromancap\romannumeral #1@}
\makeatother

\title{Joint Secure Transmit Beamforming Designs for Integrated Sensing and Communication Systems
\thanks{Part of this paper has been presented in the IEEE Wireless Communications and Networking Conference (WCNC), 2022 \cite{Chu WCNC 22}.}
\thanks{J. Chu, R. Liu, M. Li, and Y. Liu are with the School of Information and Communication Engineering, Dalian University of Technology, Dalian 116024, China (e-mail: jinjinchu@mail.dlut.edu.cn; liurang@mail.dlut.edu.cn; mli@dlut.edu.cn; yangliu\_613@dlut.edu.cn).}
\thanks{Q. Liu is with the School of Computer Science and Technology, Dalian University of Technology, Dalian 116024, China (e-mail: qianliu@dlut.edu.cn).}
}

\author{Jinjin Chu,
        Rang Liu,~\IEEEmembership{Graduate Student Member,~IEEE,}
        Ming Li,~\IEEEmembership{Senior Member,~IEEE,}\\
        Yang Liu,~\IEEEmembership{Member,~IEEE,}
        and Qian Liu,~\IEEEmembership{Member,~IEEE}
}

\begin{document}
\maketitle
\thispagestyle{empty}
\begin{abstract}
Integrated sensing and communication (ISAC), which allows individual radar and communication systems to share the same spectrum bands, is an emerging and promising technique for alleviating spectrum congestion problems.
In this paper, we investigate how to exploit the inherent interference from strong radar signals to ensure the physical layer security (PLS) for the considered multi-user multi-input single-output (MU-MISO) communication and colocated multi-input multi-output (MIMO) radar coexistence system.
In particular, with known eavesdroppers' channel state information (CSI), we propose to jointly design the transmit beamformers of communication and radar systems to minimize the maximum eavesdropping signal-to-interference-plus-noise ratio (SINR) on multiple legitimate users, while guaranteeing the communication quality-of-service (QoS) of legitimate transmissions, the requirement of radar detection performance, and the transmit power constraints of both radar and communication systems.
When eavesdroppers' CSI is unavailable, we develop a joint artificial noise (AN)-aided transmit beamforming design scheme, which utilizes residual available power to generate AN for disrupting malicious receptions as well as satisfying the requirements of both legitimate transmissions and radar target detection.
Extensive simulations verify the advantages of the proposed joint beamforming designs for ISAC systems on secure transmissions and the effectiveness of the developed algorithms.

\end{abstract}


\section{Introduction}
%\vspace{0.1 cm}
%% 引出ISAC
With the explosive growth of wireless devices, exponentially increased bandwidth is required to support a variety of high data-rate services.
Consequently, spectrum resources have been increasingly scarce, which motivates the development of advanced spectrum sharing technologies  \cite{Chiriyath TCCN 2017}.
Since large portions of the spectrum are available at radar frequency bands, spectrum sharing between radar and communication systems has led to substantial research interest \cite{Liu TCOM 2020}-\cite{A. Liu 2021}.
This line of research is referred to as integrated sensing and communication (ISAC), which is also known as joint radar-communication (JRC), joint communication and radar (JCR), joint communication and radar sensing (JCAS), etc.
ISAC allows radar and communication systems to share the same spectrum bands, which can significantly improve the spectrum efficiency and thus alleviates the spectrum congestion problem.
It is foreseeable that ISAC will become a critical enabling technology for future wireless networks, supporting various vital applications, including vehicular networks \cite{F. Liu CL 2021}, Internet of Things (IoT) \cite{Y. Cui 2021}, etc.

%% ISAC两个研究方向，DFRC简介及优缺点
Research on ISAC can be generally categorized into two main directions: Dual-functional radar-communication (DFRC) and radar and communication coexistence (RCC).
The former focuses on using the same signals transmitted from one fully-shared hardware platform to simultaneously perform communication and radar sensing functionalities \cite{F. Liu DFRC TSP 2018}-\cite{R. Liu JSTSP 2022}.
Although it has the benefits of low power consumption and smaller-size platform, the trade-off between the radar and communication functionalities requires sophisticated optimizations on the unified dual-functional waveform.
Moreover, the resulting hardware complexity significantly hinders practical applications.
On the contrary, RCC enables separately deployed communication and radar platforms to cooperatively perform their respective functions using independent transmitted signals.
Therefore, RCC is more suitable for many existing scenarios such as sharing spectrum between air-borne early warning radars and 3.5 GHz time-division duplex long-term evolution (TDD-LTE) \cite{J. H. Reed WCL 2016}, \cite{F. Hessar 2016}, between battlefield/ground surveillance and vessel traffic service (VTS) radars and WLAN networks \cite{H. Griffiths 2015}, \cite{E. Grossi TSP 2020}, etc.

%%%% RCC优点 承上启下
%% RCC的优缺点及现有研究
%RCC allows individual radar and communication systems to share the spectrum. Since they transmit independent signals, their respective performance will not be greatly sacrificed.
In RCC systems, the interference management between the non-colocated base station (BS) and radar transmitter is vital for achieving good communication and radar sensing performance.
Therefore, the cooperative design of these two systems is necessary for practical applications \cite{Zheng TSP 2018}-\cite{Qian TSP 2021}.
Meanwhile, the multi-input multi-output (MIMO) architecture has been widely deployed in both radar and communication systems to provide additional spatial degrees of freedom (DoFs) for pursuing more considerable beamforming gains.

Various signal processing techniques have been proposed to design transmit beamformers for multi-antenna BS and MIMO radar to realize efficient interference management. %and achieve satisfactory performance of both radar and communication systems.
A null space projection (NSP) method was proposed in \cite{S. Sodagari GLOBECOM 2012}, \cite{Babaei GLOBECOM 2013}, where the radar transmit beamforming is designed to project the radar signals onto the null space of the effective interference channels to eliminate the interference at the communication receivers.
However, the target echo may fall into the row space of the interference channels, resulting in degraded radar performance.
As a step further, the authors of \cite{J. A. Mahal 2017} expanded the projection space to include the subspace corresponding to the small non-zero singular values under a specified threshold, expecting to control the interference to communication systems and achieve different trade-offs between radar and communication performance.
However, these NSP-based beamforming designs significantly reduce the DoFs for optimization and cannot guarantee to meet specific radar sensing requirements.
In order to overcome these disadvantages, investigations on the joint design of the transmit beamformers for both radar and communication systems have been proposed in \cite{M. Rihan TVT 2018}, \cite{X. Liu TSP 2020}, where different performance metrics of these two systems are satisfied with controllable constraints.
In addition to the above scenarios with perfectly known channel state information (CSI), a robust beamforming design was proposed in \cite{F. Liu WCL 2017} under the assumption of imperfect CSI.

%The radar waveform and communication transmit beamformer are jointly designed aiming at minimizing the CRB of DOA estimation of radar system \cite{Z. Cheng TSP 2019}.
%The system performance of the proposed scheme in \cite{Z. Cheng TSP 2019} has a significant improvement compared with the scheme of designing only one of the systems, which verified the advantages of the joint design.
%Nevertheless, all of these works assume that channel state information (CSI) is perfectly known by the BS or the radar, which is impossible in practical scenarios.
%Thus,

As mentioned above, the existing literature on beamforming designs for RCC systems essentially aims at suppressing the interference between radar and communication systems.
While radar interference is usually deemed as the most significant harmful component to the communication system, from another perspective, it can be utilized to disrupt illegal receivers to safeguard confidential information against eavesdropping.
Utilizing interfering/jamming signals to disrupt potential eavesdroppers has been widely considered in the literature on physical layer security (PLS) \cite{M. Khandaker 2018}-\cite{M. Li}.
In \cite{M. Khandaker 2018}, constructive interference was leveraged to implement secure beamforming using a symbol-level precoding approach.
In \cite{Z. Chu}, a multi-antenna cooperative jammer was employed to assist the secure communications.
In order to confuse the eavesdropper, the authors in \cite{N. Su arxiv 2021} utilized the idea of destructive interference to push the received symbols at the eavesdropper towards the destructive region where the wrong symbol will be detected.
The authors in \cite{L. Li 2016} proposed a cooperative secure transmission scheme, in which the legitimate information and interference signals lie in different subspaces at the destination of the confidential transmission, but are aligned along the same subspace at the eavesdropper.
% The proposed scheme can achieve all the boundary points of the secrecy DoF region.

Noting that the above secure beamformer designs require the knowledge of the eavesdropper's CSI, which is not always available in practical applications.
In such cases, artificial noise (AN) technology was introduced to realize PLS \cite{W.-C. Liao}, which uses a large amount of additional energy to generate interfering signals for disrupting potential malicious receptions.
The existing communication literature generally forces the AN to be uniformly distributed onto the null space of the confidential transmission channels to disturb the eavesdropper's reception but not harm legitimate users \cite{S. Goel}.
Inspired by this concept, in addition to realizing the respective performance requirements of communication and radar systems, the residual power can be used to generate AN. %In the sequel, the legitimate transmissions are not affected while the eavesdropper will suffer from severe disruptions.
However, this null space projection design can only exploit limited spatial DoFs to generate AN, especially for a system with many users.
In recent works \cite{N. Su TWC 2021}, \cite{Y. Zhou WCSP 2019}, the AN and transmit beamforming were jointly optimized to implement secure DFRC transmissions, making full use of available DoFs for generating AN.
In addition, the authors in \cite{K. V. Mishra ICASSP 2022} proposed to enhance secure performance by deploying an RIS in DFRC systems.
However, how to ensure the physical layer security in RCC systems remains an open problem.
Instead of consuming additional power to generate AN, the authors in \cite{M. Li} proposed to exploit the inherent multi-user interference as a helpful resource by converting it to act like AN or distributed friendly jammers to improve the security performance.
Inspired by the concept of interference exploitation, the radar signal, which usually has very strong signal power, is an up-and-coming candidate as the jamming signal to enhance the physical layer security performance for the considered RCC system.

Motivated by the above findings, we investigate the PLS problem for multi-user multi-input single-output (MU-MISO) communication and colocated MIMO radar coexistence systems in this paper.
In particular, the considered RCC system includes a multi-antenna BS serving multiple single-antenna users in the presence of multiple eavesdroppers and a colocated MIMO radar attempting to detect a point-like target.
We aim to exploit the coexisted strong radar signals as inherent jamming signals to disrupt the eavesdroppers' reception\footnote{Since the radar signal that is only used to detect the target does not contain any confidential information, it will not cause security/privacy problem to the considered RCC system.
%However, in DFRC systems, the dual functional signal transmitted by the BS not only carries information, but also performs the purpose of sensing the target, which may indeed lead to the disclosure of confidential information.
%Thus, the radar security/privacy problems in DFRC systems are left for future studies.
}.
The transmit beamformers of communication and radar systems and radar receive filter are jointly designed to ensure security performance and satisfy the requirements of legitimate transmissions and radar target detection.
%guaranteeing satisfactory performance of radar target detection.
%We use the maximum eavesdropping signal-to-interference-plus-noise ratio (SINR) as a optimization metric to purse the physical layer security. %% 文献
The main contributions of this paper can be summarized as follows:
\begin{itemize}
  \item We consider the physical layer security in an ISAC system and innovatively propose to exploit the coexisted strong radar signals as inherent jamming/interfering signals to weaken the reception of potential eavesdroppers and enhance the transmission security.
      Joint secure transmit beamforming and radar receive filter designs are investigated to achieve this goal.
  \item With the knowledge of eavesdroppers' CSI, we jointly design the transmit beamforming and radar receive filter to minimize the maximum eavesdropping signal-to-interference-plus-noise ratio (SINR) on multiple legitimate users, while satisfying the quality-of-service (QoS) requirements of the legitimate users, the radar output SINR constraint, and the transmit power budgets of radar and communication systems.
      An efficient algorithm based on the block coordinate descent (BCD), fractional programming (FP), and semi-definite relaxation (SDR) methods is developed to solve the resulting non-convex optimization problem.
  \item When eavesdroppers' CSI is unavailable, we propose to jointly design the AN-aided transmit beamforming and radar receive filter to maximize the power of AN under the same constraints.
      In this case, all available transmit power of the BS and radar should be utilized to generate AN as much as possible to destruct eavesdroppers' reception.
      A double-loop BCD and SDR based algorithm is employed to convert the resulting complicated non-convex optimization problem into two more tractable sub-problems that can be alternatively solved.
  \item Extensive simulation results show that the eavesdropping SINR is generally several orders of magnitude smaller than the SINR of legitimate communication users, which verify the significant advancement of utilizing radar signals as inherent jamming/interference signals to enhance the secure transmissions for ISAC systems and the effectiveness of the proposed joint secure transmit beamforming design algorithms.
\end{itemize}

The rest of this paper is organized as follows.
Section \Rmnum{2} introduces the system model of the ISAC system in the presence of eavesdroppers and develops a joint transmit beamforming and radar receive filter design with known eavesdroppers' CSI.
Section \Rmnum{3} investigates the joint AN-aided transmit beamforming and radar receive filter design without eavesdroppers' CSI.
Simulation results are demonstrated in Section \Rmnum{4}, and finally, conclusions are provided in Section \Rmnum{5}.

\textit{Notations}: Boldface lower-case and upper-case letters indicate column vectors and matrices, respectively.
$(\cdot)^H$ and $(\cdot)^{-1}$ denote the transpose-conjugate and inverse operations, respectively.
$\mathbb{C}$ denotes the set of all complex numbers.
$| a |$, $\| \mathbf{a} \|$, and $\| \mathbf{A} \|_F$ are the magnitude of a scalar $a$, the norm of a vector $\mathbf{a}$, and the Frobenius norm of a matrix $\mathbf{A}$, respectively.
$\mathbb{E}\{\cdot\}$ represents statistical expectation.
$\text{Tr}\{\mathbf{A}\}$ and $\text{Rank}\{\mathbf{A}\}$ are the trace and rank of matrix $\mathbf{A}$, respectively.
%$\mathbf{A} \succeq 0$ states that $\mathbf{A}$ is positive semidefinite.

\begin{figure}[t]
\centering
\includegraphics[width = 3.6 in]{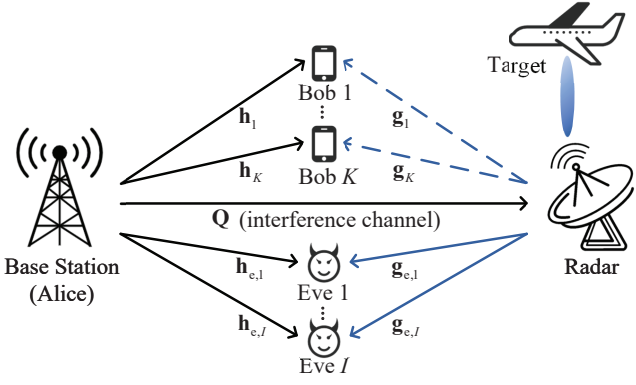}%\vspace{0.1 cm}
\caption{An ISAC system at the presence of eavesdroppers.}\label{fig:system model}%\vspace{-0.8 cm}
\end{figure}

\section{Joint Transmit Beamforming Design with Known Eavesdroppers' CSI}
%\vspace{-0.2 cm}
\subsection{System Model and Problem Formulation}
%\vspace{-0.2 cm}

We consider an ISAC system in which an MU-MISO communication system coexists with a colocated MIMO radar system operating on the same frequency band, as shown in Fig. 1.
In particular, a BS equipped with $N$ antennas in a uniform linear array (ULA) serves $K$ single-antenna users in the presence of $I$ eavesdroppers who attempt to intercept the confidential information transmissions between the BS and the legitimate users.
Meanwhile, a colocated MIMO radar with $M$ transmit/receive antennas in a ULA attempts to detect a point-like target.
For convenience, the BS, the legitimate communication users, and the eavesdroppers are referred to as Alice, Bobs, and Eves, respectively.
In this paper, we aim to exploit the coexisted strong radar signals as inherent jamming signals to disrupt eavesdroppers' reception under different assumptions about the availability of Eves' CSI.

We first assume that the CSI of Eves is perfectly known.
This assumption is valid, for example, Eves may be legitimate users who intend to overhear other users' confidential information.
With known Eves' CSI, we jointly design the transmit beamformers of the BS and the radar to ensure secure transmissions and satisfactory radar target detection performance for the ISAC system.

The received signal of the $k$-th $\text{Bob}$ can be written as
\begin{equation}
\begin{aligned}
y_k = \mathbf{h}_k^H\mathbf{W}\mathbf{s} + \mathbf{g}_k^H\mathbf{F}\mathbf{c}+ n_k,
\end{aligned}
\end{equation}
where $\mathbf{h}_k \in \mathbb{C}^{N}$ is the channel vector between the BS and the $k$-th $\text{Bob}$, $\mathbf{W}\triangleq\left[\mathbf{w}_1, \mathbf{w}_2, \ldots, \mathbf{w}_K\right] \in \mathbb{C}^{N\times K}$ is the beamforming matrix with $\mathbf{w}_k$ representing the beamforming vector of the $k$-th $\text{Bob}$, $\mathbf{s} \in \mathbb{C}^{K}$ is the symbol vector with $\mathbb{E}\{\mathbf{s}\mathbf{s}^H\}=\mathbf{I}_K$,
$\mathbf{g}_k \in \mathbb{C}^{M}$ is the channel vector between the radar and the $k$-th $\text{Bob}$,
$\mathbf{F} \in \mathbb{C}^{M\times M}$ is the radar beamforming matrix,
$\mathbf{c} \in \mathbb{C}^{M}$ is the radar transmit waveform with $\mathbb{E}\{\mathbf{c}\mathbf{c}^H\}=\mathbf{I}_M$, and $n_k$ is the additive white Gaussian noise (AWGN) with $n_k \sim \mathcal{CN}(0, \sigma_{k}^2)$.
We assume that the information symbol vector $\mathbf{s}$ is statistically independent with the radar waveform $\mathbf{c}$.
Thus, the SINR of the $k$-th $\text{Bob}$ can be calculated as
\begin{equation}
\begin{aligned}
\mathrm{SINR}_{k}&=\frac{|\mathbf{h}_k^H \mathbf{w}_k|^2 }{\sum\limits_{j \neq k}^{K}|\mathbf{h}_k^H \mathbf{w}_j|^2 + \|\mathbf{g}_k^H\mathbf{F}\|^2 + \sigma_{k}^2}\\
&=\frac{ \mathbf{h}_k^{H} \mathbf{w}_{k} \mathbf{w}_{k}^H \mathbf{h}_k}{ \sum\limits_{j \neq k}^{K} \mathbf{h}_k^{H}  \mathbf{w}_{j} \mathbf{w}_{j}^{H} \mathbf{h}_k + \mathbf{g}_k^H \mathbf{F}\mathbf{F}^H \mathbf{g}_k +\sigma_{k}^{2}}.
\end{aligned}
\end{equation}
Similarly, the received signal of the $i$-th Eve can be expressed as
\begin{equation}
\begin{aligned}
y_{\mathrm{e},i} = \mathbf{h}^H_{\mathrm{e},i} \mathbf{W}\mathbf{s} + \mathbf{g}^H_{\mathrm{e},i} \mathbf{F}\mathbf{c} + n_{\mathrm{e},i},
\end{aligned}
\end{equation}
where $\mathbf{h}_{\mathrm{e},i} \in \mathbb{C}^{N}$ is the channel vector between the BS and the $i$-th Eve,
$\mathbf{g}_{\mathrm{e},i} \in \mathbb{C}^{M}$ is the channel vector between the radar and the $i$-th Eve,
and $n_{\mathrm{e},i}$ is the AWGN with $n_{\mathrm{e},i} \sim \mathcal{CN}(0, \sigma_{\mathrm{e},i}^2)$.
The eavesdropping SINR of the $i$-th Eve on the $k$-th $\text{Bob}$ is thus given by
\begin{equation}
\begin{aligned}
\mathrm{SINR}_{i,k}^{\mathrm{e}}&=\frac{ |\mathbf{h}^H_{\mathrm{e},i}\mathbf{w}_{k}|^2 }
{\sum\limits_{j \neq k}^{K}|\mathbf{h}^H_{\mathrm{e},i}\mathbf{w}_{j}|^2+ \|\mathbf{g}^H_{\mathrm{e},i}\mathbf{F}\|^2 +\sigma_{\mathrm{e},i}^2}\\
&=\frac{ \mathbf{h}^H_{\mathrm{e},i}\mathbf{w}_{k}\mathbf{w}_{k}^{H}\mathbf{h}_{\mathrm{e},i} }
{ \sum\limits_{j \neq k}^{K} \mathbf{h}^H_{\mathrm{e},i}\mathbf{w}_{j}\mathbf{w}_{j}^{H}\mathbf{h}_{\mathrm{e},i}
+ \mathbf{g}^H_{\mathrm{e},i}\mathbf{F}\mathbf{F}^{H}\mathbf{g}_{\mathrm{e},i}
+ \sigma_{\mathrm{e},i}^2}.
\end{aligned}
\end{equation}%
To ensure secure communications, we aim to minimize the maximum eavesdropping SINR on the $K$ communication users while guaranteeing the communication QoS of legitimate transmissions.

In the radar system, the received signal, which includes the echo signal from the target, interference from the BS, and noise, can be written as
\begin{equation}
\begin{aligned}
\mathbf{y}_\mathrm{r}=\alpha\mathbf{a}_\mathrm{r}(\theta_0)\mathbf{a}_\mathrm{t}^H(\theta_0)\mathbf{F}\mathbf{c}+\mathbf{Q}^H\mathbf{W}\mathbf{s}+\mathbf{n}_\mathrm{r},
\end{aligned}
\end{equation}
where $\alpha$ is the complex target amplitude with $\mathbb{E}\{|\alpha|^2\} = \sigma_0^2$.
The vectors $\mathbf{a}_\mathrm{t}(\theta_0) \in \mathbb{C}^{M}$ and $\mathbf{a}_\mathrm{r}(\theta_0) \in \mathbb{C}^{M}$ denote the transmit and receive steering vectors of the radar antenna array,
\begin{equation}
\begin{aligned}
\mathbf{a}_\mathrm{t}(\theta_0)= \mathbf{a}_\mathrm{r}(\theta_0) \triangleq [1, e^{j\frac{2\pi}{\lambda}\Delta \mathrm{sin}(\theta_0)},  \ldots,  e^{j\frac{2\pi}{\lambda}(M-1)\Delta \mathrm{sin}(\theta_0)}]^T,
\end{aligned}
\end{equation}
where $\theta_0$ represents the azimuth angle of the target\footnote{In radar related literature, the direction of the target is typically known to the transmitter since it can be readily estimated \cite{I. Bekkerman TSP 2006}-\cite{P. Stoica TASSP 1990} at previous observations, or given by the center of angular sector-of-interest.
}, $\Delta$ denotes the antenna spacing and $\lambda$ the wavelength.
The matrix $\mathbf{Q} \in \mathbb{C}^{N \times M}$ denotes the interfering channel between the BS and the radar receiver, and $\mathbf{n}_\mathrm{r}\in \mathbb{C}^{M}$ is the AWGN with $\mathbf{n}_\mathrm{r} \sim \mathcal{CN}(0, \sigma_{\mathrm{r}}^2\mathbf{I})$.
By defining $\mathbf{A}\triangleq\mathbf{a}_\mathrm{r}(\theta_0)\mathbf{a}_\mathrm{t}^H(\theta_0)$, the received signal at the radar can be concisely re-written as
\be
\mathbf{y}_\mathrm{r}=\alpha\mathbf{A}\mathbf{F}\mathbf{c}+\mathbf{Q}^H\mathbf{W}\mathbf{s}+\mathbf{n}_\mathrm{r}.
\ee
To achieve better radar sensing performance, the radar utilizes a receive filter $\mathbf{u} \in \mathbb{C}^{M}$ to suppress the interference from the BS and the noise.
The signal after filtering is
\begin{equation}
\begin{aligned}
y_\mathrm{r}=\alpha\mathbf{u}^H\mathbf{A}\mathbf{F}\mathbf{c}+\mathbf{u}^H\mathbf{Q}^H\mathbf{W}\mathbf{s}+\mathbf{u}^H\mathbf{n}_\mathrm{r},
\end{aligned}
\end{equation}
and radar output SINR can thus be written as
\begin{equation}
\begin{aligned}
\mathrm{SINR}_\mathrm{r}
&=\frac{ \| \alpha\mathbf{u}^H\mathbf{A}\mathbf{F} \|^2 }
{  \sum\limits_{k=1}^{K}  |\mathbf{u}^H\mathbf{Q}^H\mathbf{w}_k|^2
+ \mathbb{E}\left\{|\mathbf{u}^H\mathbf{n}_\mathrm{r}|^2 \right\}  } \\
&=\frac{ \sigma_{\mathrm{0}}^2\mathbf{u}^H\mathbf{A}\mathbf{F}\mathbf{F}^H\mathbf{A}^H\mathbf{u}}
{ \mathbf{u}^H \Big(\sum\limits_{k=1}^{K}\mathbf{Q}^H\mathbf{w}_{k} \mathbf{w}_{k}^H\mathbf{Q}
+\sigma_{\mathrm{r}}^2\mathbf{I} \Big)\mathbf{u} }.
\end{aligned}
\end{equation}
For radar systems, the radar SINR is widely utilized as the metric to evaluate radar sensing performance.
Thus, we attempt to design the transmit beamforming and the radar receive filter to guarantee that the radar SINR is no less than a pre-defined threshold for achieving satisfactory radar target detection performance.

In the considered MU-MISO communication and MIMO radar coexisted system, the CSI $\mathbf{h}_k$, $\mathbf{g}_k$, $\forall k$ and $\mathbf{Q}$ are assumed to be perfectly known by appropriate channel estimation approaches\footnote{
The CSI $\mathbf{h}_k$, $\forall k$ can be obtained by conventional uplink training, i.e., users send orthogonal pilot sequences and the BS estimates CSI by classic channel estimation algorithms.
Utilizing the same pilot signals sent by users, the CSI $\mathbf{g}_k$, $\forall k$ can also be acquired by the radar without additional signaling overhead.
The channel $\mathbf{Q}$ between the BS and the radar requires specific pilot signaling for channel estimation.
Fortunately, since the geometric locations of the BS and the radar are fixed, channel $\mathbf{Q}$ is generally quasi-static and requires less estimation, which allows acceptable signaling overhead.
}.
Our objective, in this case, is to jointly design the BS transmit beamformer $\mathbf{W}$, the radar transmit beamformer $\mathbf{F}$ and the radar receive filter $\mathbf{u}$ to minimize the maximum eavesdropping SINR of the $I$ eavesdroppers on the $K$ communication users.
Meanwhile, the communication QoS requirements of the legitimate users, the radar target detection performance constraint and the power constraints of both communication and radar systems are satisfied.
Therefore, the optimization problem can be formulated as

\begin{subequations}
\label{original problem}
\begin{align}
\underset{\mathbf{W}, \mathbf{F}, \mathbf{u}}{\min} \quad & \underset{i,k}{\max} \quad \mathrm{SINR}_{i,k}^{\mathrm{e}}\\
\text { s.t. } \quad &\mathrm{SINR}_{k} \geq \Gamma_k, ~ \forall k, \\
&\mathrm{SINR}_\mathrm{r} \geq \Gamma_\mathrm{r}, \\
&\left\| \mathbf{W} \right\|_{F}^2 \leq P_\mathrm{c},\\
&\left\| \mathbf{F} \right\|_{F}^2 \leq P_\mathrm{r},
\end{align}
\end{subequations}
where $\Gamma_k$ is the SINR requirement of the $k$-th $\text{Bob}$, $\Gamma_\mathrm{r}$ is the pre-defined threshold for achieving required target detection performance, $P_\mathrm{c}$ and $P_\mathrm{r}$ denote the total power budgets of the BS and the radar, respectively.
Due to the quadratic fractional objective (\ref{original problem}a), the quadratic fractional constraints (\ref{original problem}b) and (\ref{original problem}c), and the coupled variables, problem (\ref{original problem}) is a complicated non-convex problem that cannot be directly solved.
In order to tackle these difficulties, in the next subsection, we first utilize the BCD method to convert the original problem into two sub-problems, and then employ efficient algorithms based on FP and SDR methods to iteratively solve them.

\subsection{Proposed Joint Transmit Beamforming Design}

%In this subsection, we propose an efficient BCD-FP-SDR based algorithm to tackle the non-convex optimization problem (\ref{original problem}).
%We first apply the BCD algorithm to tackle the coupling variables.
%Then employ the FP method to handle the fractional objective and the SDR method to tackle the quadratic terms in the constraints of the min-max problem.
%We describe the details of the proposed algorithm as follows.

In order to decouple the transmit beamformers $\mathbf{W}$ and $\mathbf{F}$ and the radar receive filter $\mathbf{u}$ in the non-convex constraint (\ref{original problem}c), we adopt the BCD method to iteratively solve them, which are described in detail as follows.

\textit{Update $\mathbf{W}$ and $\mathbf{F}$}: With fixed $\mathbf{u}$, the sub-problem for updating $\mathbf{W}$ and $\mathbf{F}$ is re-arranged as
\begin{subequations}
\label{eq:known update W and F}
\begin{align}
&\underset{\mathbf{w}_{k}, \forall k, \mathbf{F}}{\min} ~~  \underset{i,k}{\max} ~~ \frac{ \mathbf{h}^H_{\mathrm{e},i}\mathbf{w}_{k}\mathbf{w}_{k}^{H}\mathbf{h}_{\mathrm{e},i} }
{ \hspace{-1.5mm}\sum\limits_{j \neq k}^{K}\hspace{-1mm} \mathbf{h}^H_{\mathrm{e},i}\mathbf{w}_{j}\mathbf{w}_{j}^{H}\mathbf{h}_{\mathrm{e},i}
\hspace{-0.07cm}+\hspace{-0.07cm} \mathbf{g}^H_{\mathrm{e},i}\mathbf{F}\mathbf{F}^{H}\mathbf{g}_{\mathrm{e},i}
\hspace{-0.07cm}+\hspace{-0.07cm} \sigma_{\mathrm{e},i}^2}\\
&\text {s.t.}~~~\frac{ \mathbf{h}_k^{H} \mathbf{w}_{k} \mathbf{w}_{k}^H \mathbf{h}_k}{\sum\limits_{j \neq k}^{K} \mathbf{h}_k^{H} \mathbf{w}_{j} \mathbf{w}_{j}^{H} \mathbf{h}_k \hspace{-0.05cm}+\hspace{-0.05cm} \mathbf{g}_k^H \mathbf{F}\mathbf{F}^H\mathbf{g}_k \hspace{-0.05cm}+\hspace{-0.05cm}\sigma_{k}^{2}}  \hspace{-0.05cm}\geq \hspace{-0.05cm} \Gamma_k,\forall k, \\
&~~~~~~\frac{ \sigma_{\mathrm{0}}^2\mathbf{u}^H\mathbf{A}\mathbf{F}\mathbf{F}^H\mathbf{A}^H\mathbf{u}}
{ \mathbf{u}^H \Big(\sum\limits_{k=1}^{K}\mathbf{Q}^H\mathbf{w}_{k} \mathbf{w}_{k}^H\mathbf{Q}
+\sigma_{\mathrm{r}}^2\mathbf{I} \Big)\mathbf{u} } \geq  \Gamma_\mathrm{r}, \\
&~~~~~~\sum\limits_{k=1}^{K} \left\| \mathbf{w}_k \right\|^2 \leq P_\mathrm{c},\\
&~~~~~~\left\| \mathbf{F} \right\|_{F}^2 \leq P_\mathrm{r},
\end{align}
\end{subequations}
which is a complicated min-max problem.
To tackle the min-max problem, we introduce an auxiliary variable $z$ to re-formulate it into a more favorable form as
\begin{subequations}
\label{eq:re-formulation}
\begin{align}
&\underset{\mathbf{w}_{k}, \forall k, \mathbf{F}, z}{\min} \quad z \\
&\text {s.t.}
~~\frac{ \mathbf{h}^H_{\mathrm{e},i}\mathbf{w}_{k}\mathbf{w}_{k}^{H}\mathbf{h}_{\mathrm{e},i} }
{ \sum\limits_{j \neq k}^{K} \hspace{-0.1cm} \mathbf{h}^H_{\mathrm{e},i}\mathbf{w}_{j}\mathbf{w}_{j}^{H}\mathbf{h}_{\mathrm{e},i}
\hspace{-0.1cm}+\hspace{-0.1cm} \mathbf{g}^H_{\mathrm{e},i}\mathbf{F}\mathbf{F}^{H}\mathbf{g}_{\mathrm{e},i}
\hspace{-0.1cm}+\hspace{-0.1cm} \sigma_{\mathrm{e},i}^2}
\hspace{-0.1cm} \leq z,~  \forall i,k,\\
&\qquad(\ref{eq:known update W and F}\text{b})-(\ref{eq:known update W and F}\text{e}),
\end{align}
\end{subequations}
which is a minimization problem but still difficult to solve due to the fractional constraint (\ref{eq:re-formulation}b) and the non-convex constraints (\ref{eq:known update W and F}b) and (\ref{eq:known update W and F}c).
Noting that problem (\ref{eq:re-formulation}) has a similar form as the max-min-ratio fractional programming problems \cite{K. Shen}, Dinkelbach's transform can be applied to convert it into a more tractable form \cite{W. Dinkelbach}.
Specifically, the fractional constraint (\ref{eq:re-formulation}b) can be converted into a polynomial expression \eqref{eq:c} as shown on the top of this page
%\textcolor{blue}{
%\begin{multline}
%\label{eq:c}
%\mathbf{h}^H_{\mathrm{e},i}\mathbf{w}_{k}\mathbf{w}_{k}^{H}\mathbf{h}_{\mathrm{e},i} \\
%-c_{i,k}\big(\hspace{-3mm}\sum\limits_{j=1, j \neq k}^{K}\hspace{-3mm} \mathbf{h}^H_{\mathrm{e},i}\mathbf{w}_{j}\mathbf{w}_{j}^{H}\mathbf{h}_{\mathrm{e},i}
%+ \mathbf{g}^H_{\mathrm{e},i}\mathbf{F}\mathbf{F}^{H}\mathbf{g}_{\mathrm{e},i}
%+ \sigma_{\mathrm{e},i}^2\big)\leq t, ~\forall i,k,
%\end{multline}
%by introducing the auxiliary variable $c_{i,k}$, which essentially represents the eavesdropping SINR of the $i$-th Eve on the $k$-th Bob and is alternatively updated with the transmit beamformers $\mathbf{w}_k, \forall k$, and $\mathbf{F}$.
%}
\begin{figure*}
\begin{equation}
\begin{aligned}
\label{eq:c}
\mathbf{h}^H_{\mathrm{e},i}\mathbf{w}_{k}\mathbf{w}_{k}^{H}\mathbf{h}_{\mathrm{e},i}
-c_{i,k}\big(\sum\limits_{j \neq k}^{K} \mathbf{h}^H_{\mathrm{e},i}\mathbf{w}_{j}\mathbf{w}_{j}^{H}\mathbf{h}_{\mathrm{e},i}
+ \mathbf{g}^H_{\mathrm{e},i}\mathbf{F}\mathbf{F}^{H}\mathbf{g}_{\mathrm{e},i}
+ \sigma_{\mathrm{e},i}^2\big)\leq z, ~\forall i,k,
\end{aligned}
\end{equation}
\hrulefill
\end{figure*}
by introducing an auxiliary variable $c_{i,k}$, which essentially represents the eavesdropping SINR of the $i$-th Eve on the $k$-th Bob and is alternatively updated with the transmit beamformers $\mathbf{w}_k, \forall k$, and $\mathbf{F}$.

With given $\mathbf{w}_k, \forall k$, and $\mathbf{F}$, the optimal $c_{i,k}^\star$ can be easily obtained by
\begin{equation}
\label{eq:update c}
\begin{aligned}
c_{i,k}^\star = \frac{ \mathbf{h}^H_{\mathrm{e},i}\mathbf{w}_{k}\mathbf{w}_{k}^{H}\mathbf{h}_{\mathrm{e},i} }
{ \sum\nolimits_{j \neq k}^{K}\hspace{-0.5mm} \mathbf{h}^H_{\mathrm{e},i}\mathbf{w}_{j}\mathbf{w}_{j}^{H}\mathbf{h}_{\mathrm{e},i}
\hspace{-0.5mm}+\hspace{-0.5mm} \mathbf{g}^H_{\mathrm{e},i}\mathbf{F}\mathbf{F}^{H}\mathbf{g}_{\mathrm{e},i}
\hspace{-0.5mm}+\hspace{-0.5mm} \sigma_{\mathrm{e},i}^2}, ~\forall i,k.
\end{aligned}
\end{equation}
Then, the optimization problem for updating  $\mathbf{w}_{k}$, $\forall k$, and $\mathbf{F}$ can be formulated as
\begin{subequations}
\label{eq:known FP}
\begin{align}
&\underset{\mathbf{w}_{k}, \forall k, \mathbf{F}, z}{\min} \quad z \\ \nonumber
&\text {s.t.} ~~\eqref{eq:c} \\
&~~~~~~\frac{ \mathbf{h}_k^{H} \mathbf{w}_{k} \mathbf{w}_{k}^H \mathbf{h}_k}{\sum\limits_{j \neq k}^{K} \hspace{-0.1cm}\mathbf{h}_k^{H} \mathbf{w}_{j} \mathbf{w}_{j}^{H} \mathbf{h}_k \hspace{-0.05cm}+\hspace{-0.05cm} \mathbf{g}_k^H \mathbf{F}\mathbf{F}^H\mathbf{g}_k \hspace{-0.05cm}+\hspace{-0.05cm}\sigma_{k}^{2}}  \hspace{-0.05cm}\geq \hspace{-0.05cm} \Gamma_k,\forall k, \\
&~~~~~~\frac{ \sigma_{\mathrm{0}}^2\mathbf{u}^H\mathbf{A}\mathbf{F}\mathbf{F}^H\mathbf{A}^H\mathbf{u}}
{ \mathbf{u}^H \Big(\sum\limits_{k=1}^{K}\mathbf{Q}^H\mathbf{w}_{k} \mathbf{w}_{k}^H\mathbf{Q}
+\sigma_{\mathrm{r}}^2\mathbf{I} \Big)\mathbf{u} } \geq  \Gamma_\mathrm{r}, \\
&~~~~~~\sum\limits_{k=1}^{K} \left\| \mathbf{w}_k \right\|^2 \leq P_\mathrm{c},\\
&~~~~~~\left\| \mathbf{F} \right\|_{F}^2 \leq P_\mathrm{r}.
\end{align}
\end{subequations}
It is easy to see that the constraints \eqref{eq:c}, (\ref{eq:known FP}b), and (\ref{eq:known FP}c) are non-convex with respect to variables $\mathbf{w}_{k}, \forall k$, and $\mathbf{F}$ and are hard to tackle.
Therefore, we apply the SDR method to convert them into primary variables for an easier solution.
Specifically, by defining
\begin{equation}
\label{eq:SDR definition}
\begin{aligned}
&\mathbf{W}_{k}\triangleq\mathbf{w}_{k}\mathbf{w}_{k}^H, ~\forall k,\\
&\mathbf{R}_\mathrm{F}\triangleq\mathbf{F}\mathbf{F}^H,
\end{aligned}
\end{equation}
the quadratic terms $\mathbf{w}_{k}\mathbf{w}_{k}^H$ and $\mathbf{F}\mathbf{F}^H$ are converted into the primary variables $\mathbf{W}_{k}$ and $\mathbf{R}_\mathrm{F}$, respectively.
In the meantime, the rank-one Hermitian positive semidefinite matrices $\mathbf{W}_{k}$, $\forall k$, and the Hermitian positive semidefinite matrix $\mathbf{R}_\mathrm{F}$ should satisfy
\begin{subequations}
\label{eq:Wk constraint}
\begin{align}
&\mathbf{W}_{k}=\mathbf{W}_{k}^H, ~\mathbf{W}_{k} \succeq 0,~\forall k, \\
&\mathrm{Rank}(\mathbf{W}_{k})=1,\\
&\mathbf{R}_\mathrm{F}=\mathbf{R}_\mathrm{F}^H, ~\mathbf{R}_\mathrm{F} \succeq 0.
\end{align}
\end{subequations}
For simplicity, we define the set of all $N\times N$-dimensional Hermitian positive semidefinite matrices as $\mathbb{S}_N \triangleq \{\mathbf{S}|\mathbf{S}=\mathbf{S}^H,~\mathbf{S}\succeq 0\}$.
Afterwards, problem (\ref{eq:known FP}) is transformed into
\begin{subequations}
\label{eq:SDR}
\begin{align}
&\underset{\mathbf{W}_k, \forall k, \mathbf{R}_\mathrm{F}, z}{\min}  \quad z \\
&\text {s.t.} ~~
\mathbf{h}^H_{\mathrm{e},i}\mathbf{W}_{k}\mathbf{h}_{\mathrm{e},i} \\ \nonumber
&\quad~~~-c_{i,k}\big(\sum\limits_{j \neq k}^{K} \mathbf{h}^H_{\mathrm{e},i}\mathbf{W}_{j}\mathbf{h}_{\mathrm{e},i}
+ \mathbf{g}^H_{\mathrm{e},i}\mathbf{R}_\mathrm{F}\mathbf{g}_{\mathrm{e},i}
+ \sigma_{\mathrm{e},i}^2\big) \leq z, ~\forall i,k,\\
&\quad~~~\frac{ \mathbf{h}_k^{H} \mathbf{W}_{k} \mathbf{h}_k}{\sum\limits_{j \neq k}^{K} \mathbf{h}_k^{H} \mathbf{W}_{j} \mathbf{h}_k  + \mathbf{g}_k^H \mathbf{R}_\mathrm{F}\mathbf{g}_k +\sigma_{k}^{2}} \geq \Gamma_k,~\forall k, \\
&\quad~~~\frac{ \sigma_{\mathrm{0}}^2\mathbf{u}^H\mathbf{A}\mathbf{R}_\mathrm{F}\mathbf{A}^H\mathbf{u}}
{ \mathbf{u}^H \Big(\sum\limits_{k=1}^{K}\mathbf{Q}^H\mathbf{W}_k\mathbf{Q}
+\sigma_{\mathrm{r}}^2\mathbf{I} \Big)\mathbf{u} } \geq  \Gamma_\mathrm{r}, \\
&\quad~~~\sum_{k=1}^{K}  \mathrm{Tr}(\mathbf{W}_k)  \leq P_\mathrm{c},\\
&\quad~~~\mathrm{Tr}(\mathbf{R}_\mathrm{F}) \leq P_\mathrm{r},\\
&\quad~~~\mathbf{W}_{k}\in\mathbb{S}_N,~\forall k,~\mathbf{R}_\mathrm{F}\in\mathbb{S}_M, \\
&\quad~~~\mathrm{Rank}(\mathbf{W}_{k})=1,~\forall k.
\end{align}
\end{subequations}
It is evident that the rank-one constraint (\ref{eq:SDR}h) tremendously hinders finding a straightforward solution.
Thus, we apply the SDR algorithm by dropping the rank-one constraint (\ref{eq:SDR}h) and relaxing the problem (\ref{eq:SDR}) as
\begin{subequations}
\label{eq:SDP}
\begin{align}
\underset{\mathbf{W}_k, \forall k, \mathbf{R}_\mathrm{F}, z}{\min} ~~ &z \\
\text {s.t.} \qquad
&(\ref{eq:SDR}\text{b})-(\ref{eq:SDR}\text{g}),
\end{align}
\end{subequations}
which is a semi-definite programming (SDP) problem that can be efficiently solved by various existing algorithms and toolboxes such as CVX.
Since the rank-one constraint is temporarily neglected, the optimal objective value of problem (\ref{eq:SDP}) only serves as a lower bound.
After obtaining $\mathbf{W}_k, \forall k$, the eigenvalue decomposition (EVD) is usually used to obtain the optimal solution $\mathbf{w}_{k}$ if the resulting $\mathbf{W}_{k}$ satisfies the rank-one constraint.
Otherwise, Gaussian randomization is required to convert the high-rank solution to a feasible rank-one solution to the problem (\ref{eq:known update W and F}).
In our considered case, the rank-1 solution can be guaranteed, whose proof is given in Appendix A.
On the other hand, with resulting optimal $\mathbf{R}_\text{F}$, the radar beamforming matrix $\mathbf{F}$ can be obtained by utilizing Cholesky decomposition.

\begin{algorithm}[!t]
  \caption{Joint transmit beamforming design algorithm for solving problem (\ref{original problem})}
  \label{alg:Algorithm 1}
  \begin{algorithmic}[1]
    \REQUIRE $\mathbf{h}_{\mathrm{e},i}$, $\mathbf{g}_{\mathrm{e},i}$, $\sigma_{\mathrm{e},i}^2$, $\forall i$, $\mathbf{h}_k$, $\mathbf{g}_k$, $\sigma_k^2$, $\Gamma_k$, $\forall k$, $\mathbf{A}$, $\mathbf{Q}$, $\sigma_\text{0}^2$, $\sigma_\text{r}^2$, $~~~~~\Gamma_\mathrm{r}$, $P_\mathrm{c}$, $P_\mathrm{r}$.
    \ENSURE  $\mathbf{w}_{k}^{\star}$, $\forall k$, $\mathbf{F}^{\star}$, and $\mathbf{u}^{\star}$.
    \STATE {Initialize $\mathbf{F}$, $\mathbf{u}$, and $c_{i,k}$, $\forall i,k$.}
    \WHILE {the objective value (\ref{original problem}a) does not converge}
    \WHILE {no convergence}
    \STATE {Calculate $\mathbf{W}_{k}, \forall k$, and $\mathbf{R}_{\mathrm{F}}$ by solving (\ref{eq:SDP}).}
    \STATE {Update $\mathbf{w}_k$ from $\mathbf{W}_{k}, \forall k$, by EVD.}
    \STATE{Update $\mathbf{F}$ from $\mathbf{R}_{\mathrm{F}}$ by Cholesky decomposition.}
    \STATE {Update $c_{i,k}$, $\forall i,k$, by (\ref{eq:update c}).}
    \ENDWHILE
    \STATE {Update $\mathbf{u}$ by solving (\ref{eq:known update u}).}
    \ENDWHILE
    \STATE{Return $\mathbf{w}_k^{\star}, \forall k$, $\mathbf{F}^{\star}$, and $\mathbf{u}^{\star}$.}
  \end{algorithmic}
\end{algorithm}

\textit{Update $\mathbf{u}$}: It can be seen that the variable $\mathbf{u}$ only exists in the constraint (\ref{original problem}\text{c}) of the problem (\ref{original problem}).
Therefore, with given $\mathbf{W}$ and $\mathbf{F}$, problem (\ref{original problem}) is transformed into a feasibility-check problem.
For the sake of leaving enough freedoms for solving $\mathbf{W}$ and $\mathbf{F}$ in the subsequent optimization process and accelerating the convergence, we propose to optimize $\mathbf{u}$ to maximize the radar output SINR.
The optimization problem is formulated as
\be
\label{eq:known update u}
\mathbf{u}^\mathrm{opt}=\arg\mathop{\max}\limits_{\mathbf{u}} \frac{ \sigma_{\mathrm{0}}^2\mathbf{u}^H\mathbf{A}\mathbf{F}\mathbf{F}^H\mathbf{A}^H\mathbf{u}}
{ \mathbf{u}^H \Big(\sum\limits_{k=1}^{K}\mathbf{Q}^H\mathbf{w}_{k} \mathbf{w}_{k}^H\mathbf{Q}
+\sigma_{\mathrm{r}}^2\mathbf{I} \Big)\mathbf{u} }.
\ee
We observe that problem (\ref{eq:known update u}) is a typical generalized Rayleigh quotient, whose optimal solution can be easily obtained as the
generalized eigenvector corresponding to the largest eigenvalue of the matrix $\sigma_{\mathrm{0}}^2\big( \sum\nolimits_{k=1}^{K}\mathbf{Q}^H\mathbf{w}_{k} \mathbf{w}_{k}^H \mathbf{Q} +\sigma_{\mathrm{r}}^2\mathbf{I}\big)^{-1}\mathbf{A}\mathbf{F}\mathbf{F}^H\mathbf{A}^H$.

\subsection{Summary, Initialization, Convergence, and Complexity Analysis}

\subsubsection{Summary}
Based on the above derivations, the proposed joint transmit beamforming and radar receive filter design algorithm is straightforward and summarized in Algorithm 1.
With appropriate initialization, problems \eqref{eq:known update W and F} and \eqref{eq:known update u} are iteratively solved to respectively update $\mathbf{w}_{k}$, $\forall k$, $\mathbf{F}$, and $\mathbf{u}$ until the objective value (\ref{original problem}a) converges.
For updating $\mathbf{w}_{k}$, $\forall k$, and $\mathbf{F}$, we iteratively update the communication beamforming $\mathbf{w}_k$, $\forall k$, and the radar beamforming $\mathbf{F}$ by solving the SDP problem (\ref{eq:SDP}) and then recovering feasible solutions using EVD and Cholesky decomposition, and the auxiliary variables $c_{i,k}, \forall i, k$, until the convergence of problem (\ref{eq:known update W and F}) is found.

\subsubsection{Initialization}

In order to solve sub-problem \eqref{eq:known FP} for updating $\mathbf{w}_{k}$, $\forall k$, and $\mathbf{F}$, we need to initialize $\mathbf{u}$ and the auxiliary variables $c_{i,k},~\forall i, k$.
The initial value of the radar receive filter is selected as $\mathbf{u} = \mathbf{a}_\mathrm{r}(\theta_0)$ via a phase alignment operation for better radar detection performance.
Since $c_{i,k}$ represents the eavesdropping SINR, which is generally several orders of magnitude smaller than the communication SINR, initializing $c_{i,k}$ with the pre-defined threshold of communication SINR $\Gamma_k$ can guarantee the feasibility.
The obtained $\mathbf{w}_{k}$, $\forall k$, and $\mathbf{F}$ by solving \eqref{eq:known FP} can be set as the initial value for solving sub-problem \eqref{eq:known update u}.

\subsubsection{Convergence}

We will briefly prove the convergence of the proposed algorithm as follows.
Denote $\eta\big(\mathbf{W},\mathbf{F},\mathbf{u}\big)$ as the objective value of the original problem \eqref{original problem}.
First, in the transmit beamformer design, we apply Dinkelbach's transform to convert it into a more tractable form and transform it into the problem \eqref{eq:known FP}.
According to \cite{J.-P. Crouzeix}, it is easy to prove the convergence of the algorithm given the non-increasing property of the auxiliary variable $c_{i,k}$.
Since the optimal solution of problem \eqref{eq:known update W and F} is obtained with given $\mathbf{u}^t$, we have
\begin{equation}
\eta\big(\mathbf{W}^t,\mathbf{F}^t,\mathbf{u}^t\big)\geq
\eta\big(\mathbf{W}^{t+1},\mathbf{F}^{t+1},\mathbf{u}^t\big),
\end{equation}
where the superscript $t$ denotes the index of iterations.
Second, with fixed $\left\{\mathbf{W},\mathbf{F}\right\}$, the sub-problem for updating $\mathbf{u}$ is a feasibility-check problem.
After solving the problem \eqref{eq:known update u}, a better radar output SINR than the original requirement is achieved with the obtained radar receive filter $\mathbf{u}$ in the current iteration, i.e., the feasible domain of the original problem \eqref{original problem} is expanded while the objective value is fixed.
In other words, with given $\left\{\mathbf{W}^{t+1},\mathbf{F}^{t+1}\right\}$, we have
\begin{equation}
\eta\big(\mathbf{W}^{t+1},\mathbf{F}^{t+1},\mathbf{u}^t\big)=
\eta\big(\mathbf{W}^{t+1},\mathbf{F}^{t+1},\mathbf{u}^{t+1}\big).
\end{equation}
Based on the above analysis, we have the relationship of the objective values between iterations as
\begin{equation}
\eta\big(\mathbf{W}^{t},\mathbf{F}^{t},\mathbf{u}^t\big)\geq
\eta\big(\mathbf{W}^{t+1},\mathbf{F}^{t+1},\mathbf{u}^{t+1}\big),
\end{equation}
which indicates that the objective value of problem \eqref{original problem} is non-increasing during the iterations of Algorithm 1.
Since the objective value of problem \eqref{original problem} is greater than zero, the proposed Algorithm 1 can converge to a local optimum point.

\subsubsection{Complexity Analysis}
In this subsection, the computational complexity of Algorithm 1 is analyzed as follows.
We first analyze the computational complexity of solving for $\mathbf{w}_k$, $\forall k$, and $\mathbf{F}$.
Problem (\ref{eq:SDP}) is a convex problem with $K$ $N\times N$-dimensional and an $M \times M$-dimensional variable to be optimized, $(I+1)K+1$ second-order cone (SOC) constraints and $K+1$ linear matrix inequality (LMI) constraints.
Using the CVX solver, the computational complexity is of order $\mathcal{O}\{\text{ln}(1/\xi)2\sqrt{5}IK^{1.5}M^6\}$, where $\xi$ is the convergence threshold.
The computational complexity of updating $c_{i,k}$, $\forall i, k$, is of order $\mathcal{O}\{M^3\}$.
Other calculations have much lower complexities.
For example, updating $\mathbf{w}_k$, $\forall k$, has negligible computational of order $\mathcal{O}\{N^3\}$.
Thus, the total complexity to obtain $\mathbf{w}_k, \forall k$, and $\mathbf{F}$ is of order $\mathcal{O}\{ N_\text{FP}\text{ln}(1/\xi)2\sqrt{5}IK^{1.5}M^6\}$, where $N_\text{FP}$ is the number of iterations of the inner loop.
The computational complexity of updating $\mathbf{u}$ is of order $\mathcal{O}\{M^3\}$.
Therefore, the total computational complexity of the proposed BCD-FP-SDR algorithm is of order $\mathcal{O}\{ N_\text{tot}N_\text{FP}\text{ln}(1/\xi)2\sqrt{5}IK^{1.5}M^6 \}$,
where $N_\text{tot}$ is the number of iterations of the outer loop.

\section{Joint AN-Aided Transmit Beamforming Design without Eavesdroppers' CSI}

\subsection{System Model and Problem Formulation}

When Eves are pure passive eavesdroppers, Alice is unaware of Eves' CSI or even their existence.
In the sequel, the proposed design in the previous section cannot be adopted to ensure security performance.
In such cases, AN is a very effective method to improve the physical layer security by disrupting Eves' reception.
Specifically, in addition to transmitting confidential information or probing signals, the BS and the radar also use the available transmit power to generate AN for disturbing potential eavesdroppers as much as possible.
Therefore, the transmit beamforming and AN of both the BS and the radar are jointly designed to guarantee good communication and radar sensing performance while safeguarding the communication system against potential malicious eavesdroppers.

Based on the above descriptions, the received signal of the $k$-th $\text{Bob}$ can be written as
\begin{equation}
\begin{aligned}
y_k = \mathbf{h}_k^H(\mathbf{W}\mathbf{s}+\mathbf{z}) + \mathbf{g}_k^H(\mathbf{F}\mathbf{c}+\mathbf{v})+ n_k,
\end{aligned}
\end{equation}
where $\mathbf{z} \sim \mathcal{CN}(0, \mathbf{R}_\mathrm{z})$ and $\mathbf{v} \sim \mathcal{CN}(0, \mathbf{R}_\mathrm{v})$ are AN vectors generated by the BS and the radar, respectively.
We assume that the information symbol vector $\mathbf{s}$, the AN vector $\mathbf{z}$ generated by the BS, the radar waveform $\mathbf{c}$, and the AN vector $\mathbf{v}$ generated by the radar are statistically independent of each other.
Thus, the SINR of the $k$-th $\text{Bob}$ can be calculated as
\begin{equation}
\begin{aligned}
&\mathrm{SINR}_{k}\\
&=\hspace{-1mm}\frac{|\mathbf{h}_k^H \mathbf{w}_k|^2 }{\sum\limits_{j \neq k}^{K}|  \mathbf{h}_k^H \mathbf{w}_j|^2 \hspace{-0.5mm}+\hspace{-0.5mm}\mathbb{E}\left\{|\mathbf{h}_k^H \mathbf{z}|^2 \right\} \hspace{-0.5mm}+\hspace{-0.5mm} \|\mathbf{g}_k^H\mathbf{F}\|^2 \hspace{-0.5mm}+\hspace{-0.5mm} \mathbb{E}\left\{|\mathbf{g}_k^H \mathbf{v}|^2\right\} \hspace{-0.5mm}+\hspace{-0.5mm} \sigma_{k}^2}\\
&=\hspace{-1mm}\frac{ \mathbf{h}_k^{H} \mathbf{w}_{k} \mathbf{w}_{k}^H \mathbf{h}_k}{ \mathbf{h}_k^{H} (  \sum\limits_{j \neq k}^{K} \hspace{-1mm} \mathbf{w}_{j} \mathbf{w}_{j}^{H}\hspace{-0.5mm}+\hspace{-0.5mm}\mathbf{R}_\mathrm{z} ) \mathbf{h}_k \hspace{-0.5mm}+\hspace{-0.5mm} \mathbf{g}_k^H( \mathbf{F}\mathbf{F}^H \hspace{-0.5mm}+\hspace{-0.5mm} \mathbf{R}_\mathrm{v})\mathbf{g}_k \hspace{-0.5mm}+\hspace{-0.5mm}\sigma_{k}^{2}}.
\end{aligned}
\end{equation}
From the communication perspective, the proposed joint AN-aided transmit beamforming design aims to guarantee the communication QoS requirements of legitimate transmission while interfering with Eves as much as possible.

On the radar side, the echo wave received by the radar is expressed as
\be
\mathbf{y}_\mathrm{r}=\alpha\mathbf{A}(\mathbf{F}\mathbf{c}+\mathbf{v})+\mathbf{Q}^H(\mathbf{W}\mathbf{s}+\mathbf{z})+\mathbf{n}_\mathrm{r}.
\ee
After passing through the receive filter $\mathbf{u}$, the radar output is
\be
y_\mathrm{r}=\alpha\mathbf{u}^H\mathbf{A}(\mathbf{F}\mathbf{c}+\mathbf{v})+\mathbf{u}^H\mathbf{Q}^H(\mathbf{W}\mathbf{s}+\mathbf{z})+\mathbf{u}^H\mathbf{n}_\mathrm{r}.
\ee
The radar output SINR is thus given by
\begin{equation}
\begin{aligned}
&\mathrm{SINR}_\mathrm{r}\hspace{-1mm} \\
&=\hspace{-1mm}\frac{ \| \alpha\mathbf{u}^H\mathbf{A}\mathbf{F} \|^2 }
{ \mathbb{E}\hspace{-0.7mm}\left\{| \alpha\mathbf{u}^H\mathbf{A}\mathbf{v}|^2\right\}
\hspace{-1mm}+\hspace{-2mm} \sum\limits_{k=1}^{K} \hspace{-1mm} |\mathbf{u}^H\mathbf{Q}^H\mathbf{w}_k|^2
\hspace{-1mm}+\hspace{-1mm} \mathbb{E}\hspace{-0.7mm}\left\{|\mathbf{u}^H\mathbf{Q}^H\mathbf{z}|^2
\hspace{-1mm}+\hspace{-1mm} |\mathbf{u}^H\mathbf{n}_\mathrm{r}|^2 \right\}  }\\
&=\hspace{-1mm} \frac{ \sigma_{\mathrm{0}}^2\mathbf{u}^H\mathbf{A}\mathbf{F}\mathbf{F}^H\mathbf{A}^H\mathbf{u}}
{ \mathbf{u}^H \Big( \sigma_{\mathrm{0}}^2\mathbf{A}\mathbf{R}_\mathrm{v}\mathbf{A}^H
+ \mathbf{Q}^H(\sum\limits_{k=1}^{K}\mathbf{w}_{k} \mathbf{w}_{k}^H
+\mathbf{R}_\mathrm{z})\mathbf{Q}
+\sigma_{\mathrm{r}}^2\mathbf{I} \Big)\mathbf{u} }.
\end{aligned}
\end{equation}
From the radar perspective, the radar output SINR is guaranteed to be no less than a pre-defined threshold for achieving satisfactory target detection performance.
Meanwhile, as much power as possible is used to generate AN for interfering with Eves.

It is intuitive that higher transmission power of the confidential information poses a higher risk of being intercepted, since Eves' eavesdropping SINR is directly proportional to the transmission power.
Thus, the BS should minimize the transmit power to satisfy Bobs' QoS and use the residual power to generate AN signals.
Similarly, for the radar system, the minimum required power is allocated to generate directional signals, whose main lobe points to the direction of the target for achieving satisfactory detection performance. The huge residual power is used to generate omni-directional AN signals, which will bring excellent security performance in the presence of potential eavesdroppers.
Therefore, our objective is to jointly design the BS transmit beamformer $\mathbf{W}$, the covariance $\mathbf{R}_\mathrm{z}$ of the AN vector $\mathbf{z}$ generated by the BS, the radar transmit beamformer $\mathbf{F}$, the covariance $\mathbf{R}_\mathrm{v}$ of the AN vector $\mathbf{v}$ generated by the radar, and the radar receive filter $\mathbf{u}$ to minimize the total transmit power used by the BS and radar beamformers.
Meanwhile, the communication QoS requirements of the legitimate users, the radar output SINR constraint, and the power constraints of both communication and radar systems are satisfied.
Therefore, the optimization problem is formulated as
\begin{subequations}
\label{eq:unknown original problem}
\begin{align}
\underset{\mathbf{W}, \mathbf{R}_\mathrm{z},\mathbf{F},\mathbf{R}_\mathrm{v},\mathbf{u}}{\min} \quad &\left\| \mathbf{W} \right\|_{F}^2+\left\| \mathbf{F} \right\|_{F}^2 \\
\text {s.t.} ~~ \qquad &\mathrm{SINR}_{k} \geq \Gamma_k, ~ \forall k, \\
&\mathrm{SINR}_\mathrm{r} \geq \Gamma_\mathrm{r}, \\
&\left\| \mathbf{W} \right\|_{F}^2+\mathrm{Tr}(\mathbf{R}_\mathrm{z}) = P_\mathrm{c},\\
&\left\| \mathbf{F} \right\|_{F}^2+\mathrm{Tr}(\mathbf{R}_\mathrm{v}) = P_\mathrm{r},\\
&\mathbf{R}_\mathrm{z}\in\mathbb{S}_N,~\mathbf{R}_\mathrm{v}\in\mathbb{S}_M.
\end{align}
\end{subequations}
We observe that problem (\ref{eq:unknown original problem}) is a non-convex problem that is difficult to solve for the following two reasons.
First, the variables are intricately coupled in the constraints (\ref{eq:unknown original problem}b) and (\ref{eq:unknown original problem}c).
Second, these two constraints are quadratic and fractional.
In order to efficiently solve this problem, we employ the BCD and SDR algorithms to convert it into two more tractable sub-problems and then alternately solve each of them until convergence is achieved.

\subsection{Proposed AN-aided Joint Transmit Beamforming Design}

In this subsection, we propose an efficient double-loop BCD-SDR algorithm to tackle the non-convex problem (\ref{eq:unknown original problem}).
It can be seen that problem (\ref{eq:unknown original problem}) is very complicated due to the coupling variables in the SINR constraints (\ref{eq:unknown original problem}b) and (\ref{eq:unknown original problem}c).
To this end, we divide this problem into two sub-problems with respect to the radar and communication systems, and utilize a two-block BCD algorithm to iteratively solve them.

\subsubsection{The Sub-problem for Radar System}

Given the variables $\mathbf{w}_k$, $\forall k$, and $\mathbf{R}_\mathrm{z}$ of the communication system, the transmit beamformer $\mathbf{F}$, the covariance $\mathbf{R}_\mathrm{v}$ of the AN vector $\mathbf{v}$, and the receive filter $\mathbf{u}$ of the radar system are jointly optimized.
The problem for updating $\mathbf{F}$, $\mathbf{R}_\mathrm{v}$, and $\mathbf{u}$ can be formulated as
\begin{subequations}
\label{eq:bcd radar}
\begin{align}
&\underset{\mathbf{F},\mathbf{R}_\mathrm{v},\mathbf{u}}{\min} ~~ \left\| \mathbf{F} \right\|_{F}^2 \\
&\text {s.t.} ~~ \frac{ \mathbf{h}_k^{H} \mathbf{w}_{k} \mathbf{w}_{k}^H \mathbf{h}_k}{ \mathbf{h}_k^{H} ( \hspace{-0.5mm} \sum\limits_{j \neq k}^{K}\hspace{-1.5mm}\mathbf{w}_{j} \mathbf{w}_{j}^{H}\hspace{-1mm}+\hspace{-0.7mm}\mathbf{R}_\mathrm{z} ) \mathbf{h}_k \hspace{-0.7mm}+\hspace{-0.7mm} \mathbf{g}_k^H( \mathbf{F}\mathbf{F}^H \hspace{-1mm}+\hspace{-0.7mm} \mathbf{R}_\mathrm{v})\mathbf{g}_k \hspace{-0.7mm}+\hspace{-0.7mm}\sigma_{k}^{2}} \hspace{-1mm} \geq \hspace{-1mm} \Gamma_k, \hspace{-0.5mm} \forall k, \\
&~~~~~\frac{ \sigma_{\mathrm{0}}^2\mathbf{u}^H\mathbf{A}\mathbf{F}\mathbf{F}^H\mathbf{A}^H\mathbf{u}}
{ \mathbf{u}^H \Big( \sigma_{\mathrm{0}}^2\mathbf{A}\mathbf{R}_\mathrm{v}\mathbf{A}^H
\hspace{-0.5mm}+\hspace{-0.5mm} \mathbf{Q}^H (  \hspace{-0.2mm}\sum\limits_{k=1}^{K}\hspace{-1mm} \mathbf{w}_{k} \mathbf{w}_{k}^H
\hspace{-0.5mm}+\hspace{-0.5mm}\mathbf{R}_\mathrm{z})\mathbf{Q}
\hspace{-0.5mm}+\hspace{-0.5mm}\sigma_{\mathrm{r}}^2\mathbf{I} \Big) \hspace{-0.2mm}\mathbf{u} } \hspace{-1mm}\geq\hspace{-0.7mm} \Gamma_\mathrm{r}, \\
&~~~~~\left\| \mathbf{F} \right\|_{F}^2+\mathrm{Tr}(\mathbf{R}_\mathrm{v}) = P_\mathrm{r},\\
&~~~~~\mathbf{R}_\mathrm{v}\in\mathbb{S}_M.
\end{align}
\end{subequations}
Since the variables $\mathbf{F}$, $\mathbf{R}_\mathrm{v}$, and $\mathbf{u}$ are highly coupled in constraint (\ref{eq:bcd radar}c), which makes problem (\ref{eq:bcd radar}) very difficult to solve, we adopt a two-block BCD scheme to iteratively solve them as follows.

\textit{Update $\mathbf{F}$ and $\mathbf{R}_\mathrm{v}$}: With fixed $\mathbf{u}$, the sub-problem for updating $\mathbf{F}$ and $\mathbf{R}_\mathrm{v}$ is re-arranged as
\begin{subequations}
\label{eq:update F and Rv}
\begin{align}
\underset{\mathbf{F},\mathbf{R}_\mathrm{v}}{\min} ~~ &\left\| \mathbf{F} \right\|_{F}^2 \\
\text { s.t. } \quad &(\ref{eq:bcd radar}\text{b})-(\ref{eq:bcd radar}\text{e}).
\end{align}
\end{subequations}
Considering that the constraints (\ref{eq:bcd radar}\text{b}) and (\ref{eq:bcd radar}\text{c}) are still non-convex due to the quadratic terms with respect to $\mathbf{F}$, we transform problem (\ref{eq:update F and Rv}) into
\begin{subequations}
\label{eq:calculate Rf and Rv}
\begin{align}
&\underset{\mathbf{R}_\mathrm{F},\mathbf{R}_\mathrm{v}}{\min} ~~ \mathrm{Tr}(\mathbf{R}_\mathrm{F}) \\
&\text {s.t.} ~ \frac{ \mathbf{h}_k^{H} \mathbf{w}_{k} \mathbf{w}_{k}^H \mathbf{h}_k}{ \mathbf{h}_k^{H} ( \hspace{-0.5mm} \sum\limits_{j \neq k}^{K}\hspace{-1mm}\mathbf{w}_{j} \mathbf{w}_{j}^{H} \hspace{-0.5mm}+\hspace{-0.5mm}\mathbf{R}_\mathrm{z} ) \mathbf{h}_k \hspace{-0.5mm}+\hspace{-0.5mm} \mathbf{g}_k^H( \mathbf{R}_\mathrm{F} \hspace{-0.5mm}+\hspace{-1mm} \mathbf{R}_\mathrm{v})\mathbf{g}_k \hspace{-0.5mm}+\hspace{-0.5mm}\sigma_{k}^{2}} \hspace{-1mm} \geq \hspace{-0.7mm} \Gamma_k, \hspace{-0.5mm} \forall k, \\
&\frac{ \sigma_{\mathrm{0}}^2\mathbf{u}^H\mathbf{A}\mathbf{R}_\mathrm{F}\mathbf{A}^H\mathbf{u}}{ \mathbf{u}^H \hspace{-0.5mm} \Big( \hspace{-0.7mm}\sigma_{\mathrm{0}}^2\mathbf{A}\mathbf{R}_\mathrm{v}\mathbf{A}^H \hspace{-1.4mm}+\hspace{-1mm} \mathbf{Q}^H \hspace{-0.3mm}( \hspace{-0.8mm}\sum\limits_{k=1}^{K}\hspace{-1mm}\mathbf{w}_{k} \mathbf{w}_{k}^H \hspace{-1.2mm}+\hspace{-1.2mm} \mathbf{R}_\mathrm{z})\mathbf{Q} \hspace{-0.7mm}+\hspace{-0.7mm} \sigma_{\mathrm{r}}^2\mathbf{I} \hspace{-0.3mm} \Big)\mathbf{u} } \hspace{-0.7mm}\geq\hspace{-0.7mm} \Gamma_\mathrm{r}, \\
&\mathrm{Tr}(\mathbf{R}_\mathrm{F}) + \mathrm{Tr}(\mathbf{R}_\mathrm{v}) = P_\mathrm{r},\\
&\mathbf{R}_\mathrm{F}\in\mathbb{S}_M, ~\mathbf{R}_\mathrm{v}\in\mathbb{S}_M.
\end{align}
\end{subequations}
Obviously, problem (\ref{eq:calculate Rf and Rv}) is an SDP problem and can be solved by convex tools, e.g., CVX.
After solving $\mathbf{R}_\text{F}$, the radar beamforming matrix $\mathbf{F}$ can be obtained by utilizing Cholesky decomposition.

\textit{Update $\mathbf{u}$}: Notice that the variable $\mathbf{u}$ only exists in the constraint (\ref{eq:bcd radar}\text{c}) of the problem (\ref{eq:bcd radar}).
Therefore, with given $\mathbf{F}$ and $\mathbf{R}_\mathrm{v}$, problem (\ref{eq:bcd radar}) is transformed into a feasibility-check problem.
In order to leave more freedoms for minimizing $\left\| \mathbf{F} \right\|_{F}^2$ in the next iteration, we propose to optimize $\mathbf{u}$ to maximize the radar output SINR.
The optimization problem is formulated as
\begin{equation}
\begin{aligned}
\label{eq:update u}
&\mathbf{u}^\mathrm{opt}=\\
&\arg\mathop{\max}\limits_{\mathbf{u}} \frac{ \sigma_{\mathrm{0}}^2\mathbf{u}^H\mathbf{A}\mathbf{F}\mathbf{F}^H\mathbf{A}^H\mathbf{u}}
{ \mathbf{u}^H \hspace{-0.7mm}\Big( \hspace{-0.7mm}\sigma_{\mathrm{0}}^2\mathbf{A}\mathbf{R}_\mathrm{v}\mathbf{A}^H
\hspace{-0.7mm}+\hspace{-0.7mm} \mathbf{Q}^H(\sum\limits_{k=1}^{K}\mathbf{w}_{k} \mathbf{w}_{k}^H
\hspace{-0.7mm}+\hspace{-0.7mm}\mathbf{R}_\mathrm{z})\mathbf{Q}
\hspace{-0.7mm}+\hspace{-0.7mm}\sigma_{\mathrm{r}}^2\mathbf{I} \Big)\mathbf{u} },
\end{aligned}
\end{equation}
whose optimal solution can be easily obtained as the generalized eigenvector corresponding to the largest eigenvalue of matrix $\sigma_{\mathrm{0}}^2\big[\sigma_{\mathrm{0}}^2\mathbf{A}\mathbf{R}_\mathrm{v}\mathbf{A}^H \!+ \!\mathbf{Q}^H(\sum\limits_{k=1}^{K}\mathbf{w}_{k} \mathbf{w}_{k}^H \!+\!\mathbf{R}_\mathrm{z})\mathbf{Q} +\sigma_{\mathrm{r}}^2\mathbf{I}\big]^{-1}\mathbf{A}\mathbf{F}\mathbf{F}^H\mathbf{A}^H$.

\begin{algorithm}[!t]
  \caption{Joint AN-aided transmit beamforming design algorithm for solving problem (\ref{eq:unknown original problem})}
  \label{alg:Algorithm 1}
  \begin{algorithmic}[1]
    \REQUIRE $\mathbf{h}_k$, $\mathbf{g}_k$, $\Gamma_k$, $\sigma_k^2$, $\forall k$, $\mathbf{A}$, $\mathbf{Q}$, $\sigma_{\mathrm{0}}^2$, $\sigma_\text{r}^2$, $\Gamma_\mathrm{r}$, $P_\mathrm{c}$, $P_\mathrm{r}$.
    \ENSURE  $\mathbf{w}_{k}^{\star}$, $\forall k$, $\mathbf{R}_\mathrm{z}^{\star}$, $\mathbf{F}^{\star}$, $\mathbf{R}_\mathrm{v}^{\star}$, and $\mathbf{u}^{\star}$.
    \STATE {Initialize $\mathbf{w}_{k}$, $\forall k$, $\mathbf{R}_\mathrm{z}$,  and $\mathbf{u}$.}
    \WHILE {the objective value (\ref{eq:unknown original problem}a) does not converge}
    \WHILE {the objective value (\ref{eq:bcd radar}a) does not converge}
    \STATE {Calculate $\mathbf{R}_{\mathrm{F}}$ and update $\mathbf{R}_\mathrm{v}$ by solving (\ref{eq:calculate Rf and Rv}).}
    \STATE {Update $\mathbf{F}$ from $\mathbf{R}_{\mathrm{F}}$ by Cholesky decomposition.}
    \STATE {Update $\mathbf{u}$ by (\ref{eq:update u}).}
    \ENDWHILE
    \STATE{Calculate $\mathbf{W}_{k}$, $\forall k$, and update $\mathbf{R}_\mathrm{z}$ by solving (\ref{eq:calculate Wk and Rz}).}
    \STATE {Update $\mathbf{w}_k$ from $\mathbf{W}_{k}$, $\forall k$, by EVD.}
    \ENDWHILE
    \STATE{Return $\mathbf{w}_{k}^{\star}$, $\forall k$, $\mathbf{R}_\mathrm{z}^{\star}$, $\mathbf{F}^{\star}$, $\mathbf{R}_\mathrm{v}^{\star}$, and $\mathbf{u}^{\star}$.}
  \end{algorithmic}
\end{algorithm}

\subsubsection{The Sub-problem for Communication System}

With fixed $\mathbf{F}$, $\mathbf{R}_\mathrm{v}$, and $\mathbf{u}$, the sub-problem for updating the transmit beamformer $\mathbf{W}$ and the covariance $\mathbf{R}_\mathrm{z}$ of the AN vector $\mathbf{z}$ is re-arranged as
\begin{subequations}
\label{eq:bcd communication}
\begin{align}
\underset{\mathbf{w}_k, \forall k, \mathbf{R}_\mathrm{z}}{\min} ~~ &\sum_{k=1}^{K}  \mathrm{Tr}(\mathbf{w}_k\mathbf{w}_k^H) \\
\text{s.t.} ~~\quad &(\ref{eq:bcd radar}\text{b}), ~(\ref{eq:bcd radar}\text{c})\\
&\sum_{k=1}^{K}  \mathrm{Tr}(\mathbf{w}_k\mathbf{w}_k^H)+\mathrm{Tr}(\mathbf{R}_\mathrm{z}) = P_\mathrm{c},\\
&\mathbf{R}_\mathrm{z}\in\mathbb{S}_N.
\end{align}
\end{subequations}
Similarly, we utilize the SDR algorithm to convert this problem into an SDP problem by using the definitions in (\ref{eq:SDR definition}) and dropping the rank-one constraint (\ref{eq:Wk constraint}b) as
\begin{subequations}
\label{eq:calculate Wk and Rz}
\begin{align}
&\underset{\mathbf{W}_k, \forall k, \mathbf{R}_\mathrm{z}}{\min} ~~ \sum_{k=1}^{K}  \mathrm{Tr}(\mathbf{W}_k) \\
\text {s.t.} ~&\frac{ \mathbf{h}_k^{H} \mathbf{W}_{k} \mathbf{h}_k}{ \mathbf{h}_k^{H} ( \sum\limits_{j \neq k}^{K}\hspace{-0mm}\mathbf{W}_{j}\hspace{-0.5mm}+\hspace{-0.5mm}\mathbf{R}_\mathrm{z} ) \mathbf{h}_k \hspace{-0.5mm}+\hspace{-0.5mm} \mathbf{g}_k^H( \mathbf{F}\mathbf{F}^H \hspace{-0.5mm}+\hspace{-0.5mm} \mathbf{R}_\mathrm{v})\mathbf{g}_k \hspace{-0.5mm}+\hspace{-0.5mm}\sigma_{k}^{2}} \hspace{-0mm} \geq \hspace{-0mm} \Gamma_k, \hspace{-0mm} \forall k, \\
&\frac{\sigma_{\mathrm{0}}^2\mathbf{u}^H\mathbf{A}\mathbf{F}\mathbf{F}^H\mathbf{A}^H\mathbf{u}}
{ \mathbf{u}^H \hspace{-0.0mm}\Big( \hspace{-0.0mm} \sigma_{\mathrm{0}}^2\mathbf{A}\mathbf{R}_\mathrm{v}\mathbf{A}^H
\hspace{-0.5mm}+\hspace{-0.5mm} \mathbf{Q}^H (\hspace{-0mm}\sum\limits_{k=1}^{K}\hspace{-0mm}\mathbf{W}_{k}
\hspace{-0.5mm}+\hspace{-0.5mm}\mathbf{R}_\mathrm{z})\mathbf{Q}
\hspace{-0.5mm}+\hspace{-0.5mm}\sigma_{\mathrm{r}}^2\mathbf{I} \hspace{-0mm} \Big)\mathbf{u} } \hspace{-0mm}\geq\hspace{-0mm} \Gamma_\mathrm{r}, \\
&\sum_{k=1}^{K}\mathrm{Tr}(\mathbf{W}_k)+\mathrm{Tr}(\mathbf{R}_\mathrm{z}) = P_\mathrm{c},\\
&(\ref{eq:Wk constraint}\text{a}),~(\ref{eq:bcd communication}\text{d}),
\end{align}
\end{subequations}
and then solve the resulting SDP problem by various existing algorithms or toolboxes, e.g., CVX.
The obtained $\mathbf{W}_{k}$, $\forall k$ can also be guaranteed to satisfy the rank-1 constraint in this case.
The proof is given in Appendix A.
We use the same method as in the previous section to obtain $\mathbf{w}_{k}$, $\forall k$.

\subsection{Summary, Initialization, Convergence, and Complexity Analysis}

\subsubsection{Summary}

Based on the above derivations, the proposed joint AN-aided transmit beamforming design is straightforward and summarized in Algorithm 2.
In the inner loop, we iteratively solve problems (\ref{eq:update F and Rv}) and (\ref{eq:update u}) to respectively update $\mathbf{F}$, $\mathbf{R}_\mathrm{v}$, and $\mathbf{u}$ until the objective value (\ref{eq:bcd radar}a) converges.
In the outer loop, we iteratively solve problems (\ref{eq:bcd radar}) and (\ref{eq:bcd communication}) for updating $\mathbf{w}_k$, $\forall k$, and $\mathbf{R}_\text{z}$ until the objective value (\ref{eq:unknown original problem}a) converges.

\subsubsection{Initialization}

In order to solve sub-problem \eqref{eq:update F and Rv} for updating $\mathbf{F}$ and $\mathbf{R}_\mathrm{v}$, we investigate to properly initialize $\mathbf{W}$, $\mathbf{R}_\mathrm{z}$, and $\mathbf{u}$.
The concept of NSP is utilized to initialize $\mathbf{W}$ and $\mathbf{R}_\mathrm{z}$.
Specifically, we jointly design the beamforming without AN aiming to guarantee the performance requirements of communication and radar systems, and then project the AN of the communication system onto the null space of the effective interference channels between the BS and Bobs.
Similarly, the initial value of the radar receive filter is selected as $\mathbf{u} = \mathbf{a}_\mathrm{r}(\theta_0)$ via a phase alignment operation.
The obtained $\mathbf{F}$, $\mathbf{R}_\mathrm{v}$, and $\mathbf{u}$ by solving \eqref{eq:bcd radar} can be set as the initial value for solving sub-problem \eqref{eq:bcd communication}.

\subsubsection{Convergence}
The convergence of the proposed algorithm will be briefly proven as follows.
Denote $\eta\big(\mathbf{W}, \mathbf{R}_\mathrm{z},\mathbf{F},\mathbf{R}_\mathrm{v},\mathbf{u}\big)$ as the objective value of the original problem \eqref{eq:unknown original problem}.
First, for the radar system, the sub-problem for updating $\mathbf{F}$ and $\mathbf{R}_\mathrm{v}$ is transformed into an SDP problem.
It is obvious that the transmit power is non-increasing between iterations, we have
\begin{equation}
\eta\big(\mathbf{W}^t, \mathbf{R}_\mathrm{z}^t,\mathbf{F}^t,\mathbf{R}_\mathrm{v}^t,\mathbf{u}^t\big)\geq
\eta\big(\mathbf{W}^t, \mathbf{R}_\mathrm{z}^t,\mathbf{F}^{t+1},\mathbf{R}_\mathrm{v}^{t+1},\mathbf{u}^t\big),
\end{equation}
with fixed $\left\{\mathbf{W}^{t},\mathbf{R}_\mathrm{z}^t,\mathbf{u}^t\right\}$.
Given $\left\{\mathbf{W}^{t},\mathbf{R}_\mathrm{z}^{t},\mathbf{F}^{t+1},\mathbf{R}_\mathrm{v}^{t+1}\right\}$, the sub-problem for updating $\mathbf{u}$ is a feasibility-check problem, which follows that
\begin{equation}
\eta\big(\mathbf{W}^t, \mathbf{R}_\mathrm{z}^t,\mathbf{F}^{t+1},\mathbf{R}_\mathrm{v}^{t+1},\mathbf{u}^t\big)=
\eta\big(\mathbf{W}^t, \mathbf{R}_\mathrm{z}^t,\mathbf{F}^{t+1},\mathbf{R}_\mathrm{v}^{t+1},\mathbf{u}^{t+1}\big).
\end{equation}
Second, the sub-problem for the communication system is also transformed into an SDP problem.
With given $\left\{\mathbf{F}^{t+1},\mathbf{R}_\mathrm{v}^{t+1},\mathbf{u}^{t+1}\right\}$, it follows that
\begin{equation}
\begin{aligned}
&\eta\big(\mathbf{W}^t, \mathbf{R}_\mathrm{z}^t,\mathbf{F}^{t+1},\mathbf{R}_\mathrm{v}^{t+1},\mathbf{u}^{t+1}\big)\\
&\geq
\eta\big(\mathbf{W}^{t+1}, \mathbf{R}_\mathrm{z}^{t+1},\mathbf{F}^{t+1},\mathbf{R}_\mathrm{v}^{t+1},\mathbf{u}^{t+1}\big).
\end{aligned}
\end{equation}
Based on the above analysis, we have the relationship of the objective values between iterations as
\begin{equation}
\begin{aligned}
&\eta\big(\mathbf{W}^t, \mathbf{R}_\mathrm{z}^t,\mathbf{F}^{t},\mathbf{R}_\mathrm{v}^{t},\mathbf{u}^t\big)\\
&\geq
\eta\big(\mathbf{W}^{t+1}, \mathbf{R}_\mathrm{z}^{t+1},\mathbf{F}^{t+1},\mathbf{R}_\mathrm{v}^{t+1},\mathbf{u}^{t+1}\big),
\end{aligned}
\end{equation}
which indicates that the objective value of problem \eqref{eq:unknown original problem} is non-increasing during iterations of Algorithm 2.
Since the objective value of problem \eqref{eq:unknown original problem} is greater than zero, the proposed Algorithm 2 can converge to a local optimum point.

\subsubsection{Complexity Analysis}
In the inner loop, problem (\ref{eq:calculate Rf and Rv}) is a convex problem with two $M \times M$-dimensional variables to be optimized, $K+1$ SOC constraints and an LMI constraint.
Using the CVX solver, the complexity for updating $\mathbf{F}$ and $\mathbf{R}_\mathrm{v}$ is of order $\mathcal{O}\{\text{ln}(1/\xi)8\sqrt{2}K^{1.5}M^6\}$, and solving problem (\ref{eq:update u}) for updating $\mathbf{u}$ has the complexity of order $\mathcal{O}\{M^3\}$.
Thus, the total complexity for obtaining $\mathbf{F}$, $\mathbf{R}_\mathrm{v}$, and $\mathbf{u}$ is of order $\mathcal{O}\{N_\text{inn}\text{ln}(1/\xi)8\sqrt{2}K^{1.5}M^6\}$, where $N_\text{inn}$ is the number of iterations of the inner loop.
Similarly, problem (\ref{eq:calculate Wk and Rz}) is a convex problem with $K+1$ $N \times N$-dimensional variables to be optimized, $K+1$ SOC constraints, and an LMI constraint.
The computational complexity for updating $\mathbf{w}_k$, $\forall k$, and $\mathbf{R}_\text{z}$ is of order $\mathcal{O}\{\text{ln}(1/\xi)\sqrt{2}K^{4.5}N^6\}$.
Other lower complexity calculations are omitted.
Thus, the total computational complexity for solving the problem (\ref{eq:unknown original problem}) is of order $\mathcal{O}\{N_\text{out}N_\text{inn}\text{ln}(1/\xi)8\sqrt{2}K^{1.5}M^6\}$,
where $N_\text{out}$ is the number of iterations of the outer loop.

\section{Simulation Results}

In this section, extensive simulation results are provided to demonstrate the performance of our proposed joint secure beamforming designs for the considered ISAC systems under the assumption that Eves' CSI is available or not.
Except for the radar-target link adopting the AoA model, the Rayleigh fading channel model is adopted for all links, i.e., each entry of the channel matrices is assumed to obey the standard complex Gaussian distribution.
The number of eavesdroppers is $I$ = 2.
The antenna spacing of the radar is $\Delta=\lambda/2$.
The noise power at Eves and Bobs are set as $\sigma_{\mathrm{e},i}^{2}=\sigma_{k}^{2} = 10\mathrm{dBm}, ~\forall i,k$.
The total power budget of the BS is $P_\mathrm{c}$ = 10W.
For simplicity, we assume the communication QoS requirement of each Bob is the same and denoted by $\Gamma_\mathrm{c}$.
The target is located at the azimuth angle $\theta_0=0^\circ$ and the radar cross section (RCS) is $\sigma_{\mathrm{0}}^2=1$.
The noise power to radar echoes is set as $\sigma_\mathrm{r}^{2}=0\mathrm{dB}$\footnote{The radar echo passes through the round-trip distance between the radar and the target, while the signal the user receives only passes through the one-way distance from the BS to the user. Thus, the noise power to radar echoes is usually a little larger than that of the users.}.
Besides, the SINR requirement thresholds should be set appropriately to prevent an infeasible case.

%% iter
\begin{figure}[t]
\centering
\includegraphics[width = 3.6 in]{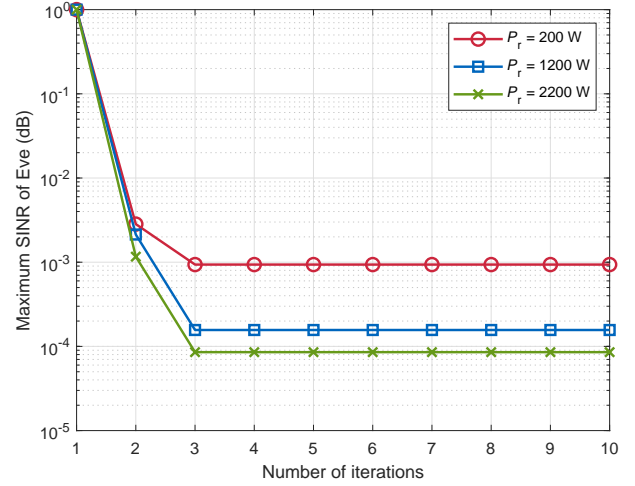}%\vspace{-0.4 cm}
\caption{Convergence of the proposed algorithm.}\label{fig:iteration} %\vspace{-0.3 cm}
\end{figure}

%% \gamma
\begin{figure}[t]
\centering
\includegraphics[width = 3.6 in]{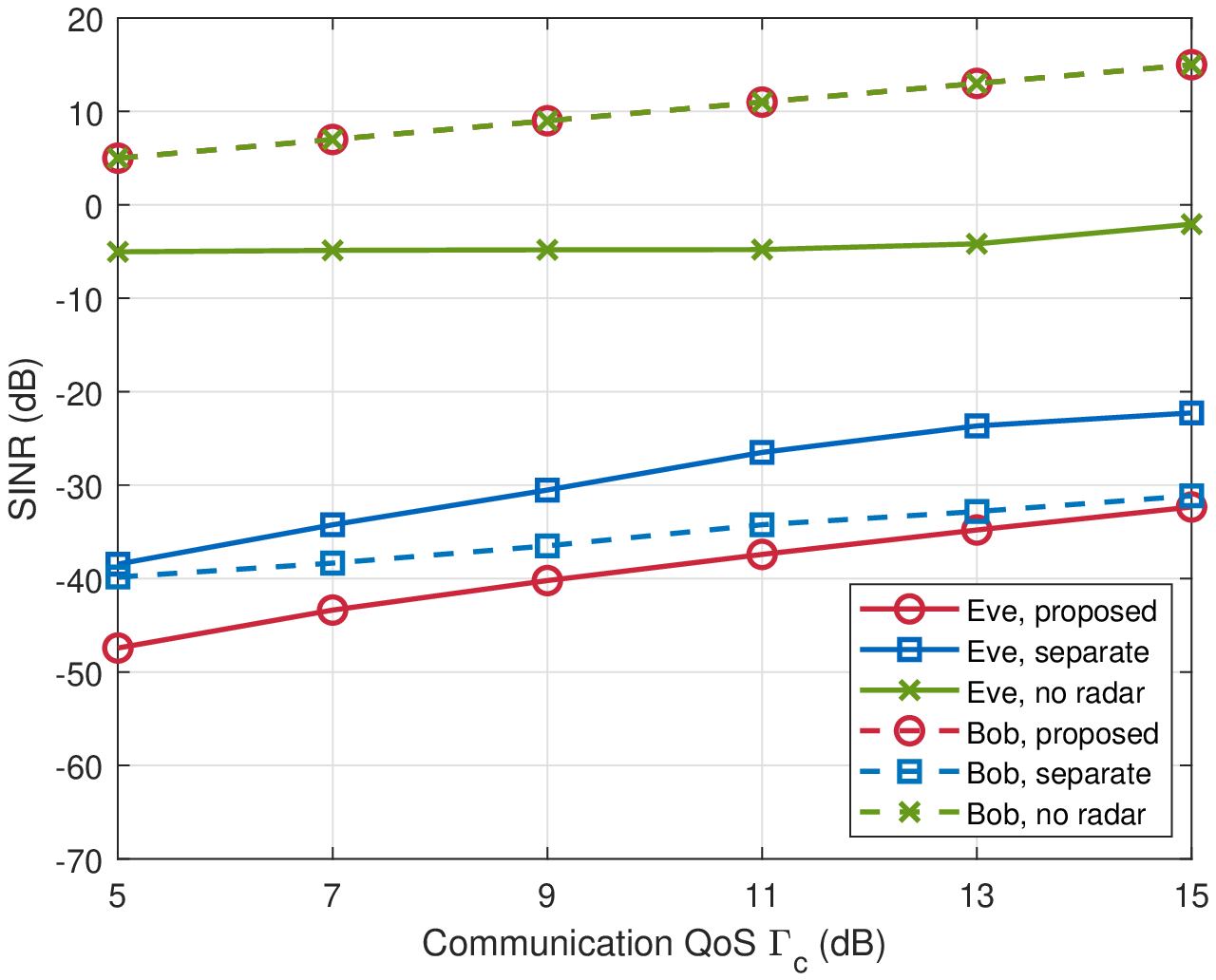}%\vspace{-0.4 cm}
\caption{SINRs of Eve and Bob versus the communication QoS requirement $\Gamma_\mathrm{c}$ ($M$ = 16, $\Gamma_\mathrm{r}$ = 10dB, $P_\mathrm{r}$ = 500W).}\label{fig:gamma} %\vspace{-0.0 cm}
\end{figure}

%% M
\begin{figure}[t]
\centering
\includegraphics[width = 3.6 in]{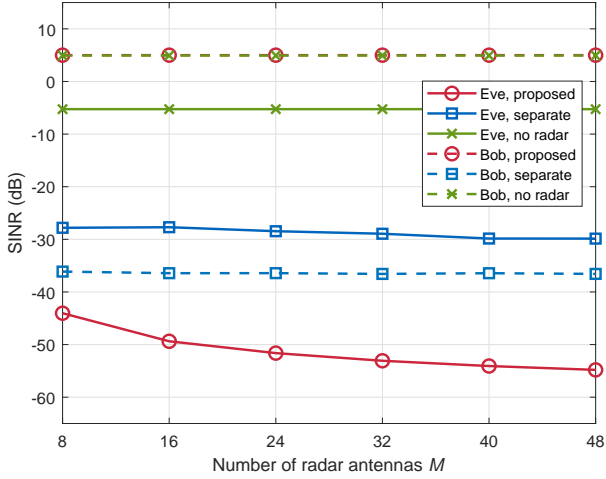}%\vspace{-0.4 cm}
\caption{SINRs of Eve and Bob versus the number of radar transmit/receive antennas $M$ ($\Gamma_\mathrm{c}$ = 5dB, $\Gamma_\mathrm{r}$ = 10dB, $P_\mathrm{r}$ = 500W).}\label{fig:M}
%\vspace{-0.2 cm}
\end{figure}

%% Pr
\begin{figure}[t]
\centering
\includegraphics[width = 3.6 in]{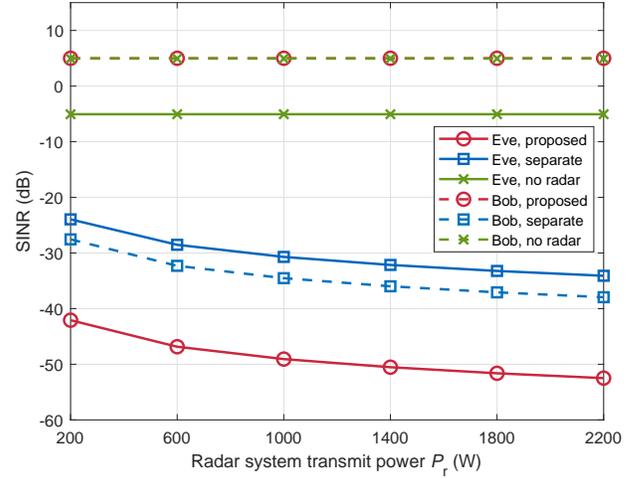}%\vspace{-0.4 cm}
\caption{SINRs of Eve and Bob versus the radar system transmit power $P_\mathrm{r}$ ($M$ = 16, $\Gamma_\mathrm{c}$ = 5dB, $\Gamma_\mathrm{r}$ = 10dB).}\label{fig:Pr}
%\vspace{-0.3 cm}
\end{figure}

\subsection{Known Eavesdroppers' CSI}

In this subsection, we illustrate the performance of our proposed joint secure beamforming design assuming perfect knowledge of Eves' CSI.
We assume that the BS is equipped with $N = 4$ antennas to serve $K = 4$ Bobs.
In all simulations, the SINR of Eve denotes the maximum eavesdropping SINR and the SINR of Bob denotes the minimum legitimate SINR, respectively.

% iteration
We first illustrate the convergence performance of the proposed algorithm in Fig. \ref{fig:iteration}.
It can be observed that our proposed algorithm converges very quickly under different settings, which reveals the effectiveness of the proposed algorithm and its great potential in practical applications.

% SINR versus QoS
The SINRs of Eve and Bob versus the communication QoS requirement $\Gamma_\mathrm{c}$ are plotted in Fig. \ref{fig:gamma} to illustrate the performance of secure communication.
The ``proposed'' scheme denotes our proposed joint transmit beamforming design for secure ISAC systems.
For comparisons, we also include the ``no radar'' scheme, which denotes that there is one single communication system without radar interference, and the ``separate'' scheme, which represents that the transmit beamformers of the communication system and radar system are separately designed without considering the interference between them.
From Fig. \ref{fig:gamma}, we can observe that in the scenario of ``no radar'', the required communication QoS of Bobs can be guaranteed, but the security of the communication system is not very satisfactory since Eves' SINR is not low enough.
When the radar joins the system without cooperative joint beamforming design, Eves' SINR is significantly reduced owing to the strong radar interfering signals, which provides good confidentiality.
However, at the same time, legitimate transmissions are also severely damaged by radar interference.
Compared with these two scenarios, our proposed joint transmit beamforming design significantly decreases Eves' SINR to a minimum value, while always maintaining Bobs' SINR at a required level to satisfy the QoS requirements of the legitimate transmissions.

% SINR versus M
In Fig. \ref{fig:M}, we show the SINRs of Eve and Bob versus the number of radar antennas $M$.
It is natural that the number of radar antennas does not affect the security and legitimate transmission in the ``no radar'' and ``separate'' scenarios.
With the proposed joint beamforming design algorithm, Eves' SINR decreases with increasing $M$ since additional spatial DoFs can be exploited to provide stronger interference to Eve.
This phenomenon verifies that a higher DoF of MIMO radar is of great help for secure communication.

% SINR versus Pr
Next, we present the SINRs of Eve and Bob versus the radar system power budget $P_\mathrm{r}$ in Fig. \ref{fig:Pr}, where the same relationship between different scenarios can be observed as that in Figs. \ref{fig:gamma} and \ref{fig:M}.
In addition, we observe that the eavesdropping SINRs of both the ``proposed'' and ``separate'' schemes decrease with a larger $P_\mathrm{r}$ due to stronger jamming signals.
Moreover, the proposed scheme achieves the least eavesdropping SINR thanks to the cooperative joint beamforming design, which guarantees the most favorable security performance.

\subsection{Unknown Eavesdroppers' CSI}

%\begin{figure}[!t]
%\centering
%\subfigure[Inner loop.]{
%\begin{minipage}{4  cm}
%\centering
%\includegraphics[width=1.75 in]{inner.eps}
%\vspace{-0.2 cm}
%\label{fig:inner loop}
%\end{minipage}}
%\subfigure[Outer loop.]{
%\begin{minipage}{4 cm}
%\centering
%\includegraphics[width=1.75 in]{outer_M16_Pc60_Pr1000_gamma10_ep10.eps}
%\vspace{-0.2 cm}
%\label{fig:outer loop}
%\end{minipage}}
%\caption{Convergence of proposed double-loop BCD-SDR algorithm.}
%\label{fig:convergence}
%\end{figure}

%% convergence
\begin{figure}[t]
\centering
\includegraphics[width = 3.6 in]{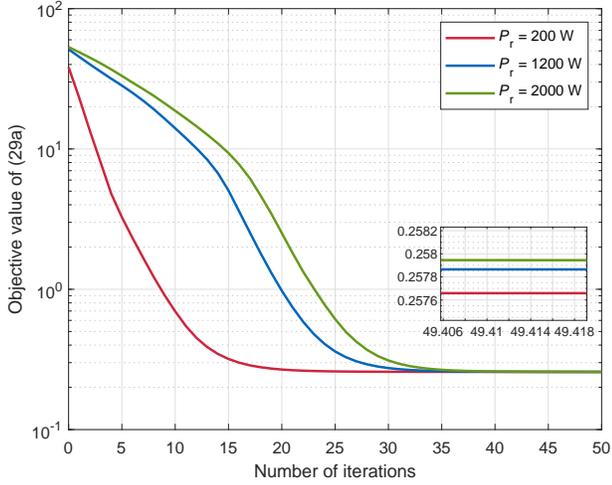}%\vspace{-0.4 cm}
\caption{Convergence of the proposed algorithm.}\label{fig:convergence} \vspace{-0.0 cm}
\end{figure}

In this subsection, we demonstrate the performance of our proposed joint AN-aided transmit beamforming design for the case that Eves' CSI is unknown.
The BS is equipped with $N=8$ antennas.
We first illustrate the convergence performance of the proposed algorithm in Fig. \ref{fig:convergence}.
%The convergence of the objective value (\ref{eq:bcd radar}a) in the inner loop is shown in Fig. \ref{fig:convergence} (a), where the curve with square markers denotes the initial loop.
%It can be seen that the inner loop quickly converges within 2 iterations.
%We can see that the convergence of the objective value (\ref{eq:unknown original problem}a),
It can be observed that the objective value (\ref{eq:unknown original problem}a) monotonically converges within a limited number of iterations under different settings, which verifies the effectiveness of the proposed algorithm.

\begin{figure}[t]
\centering
\includegraphics[width = 3.6 in]{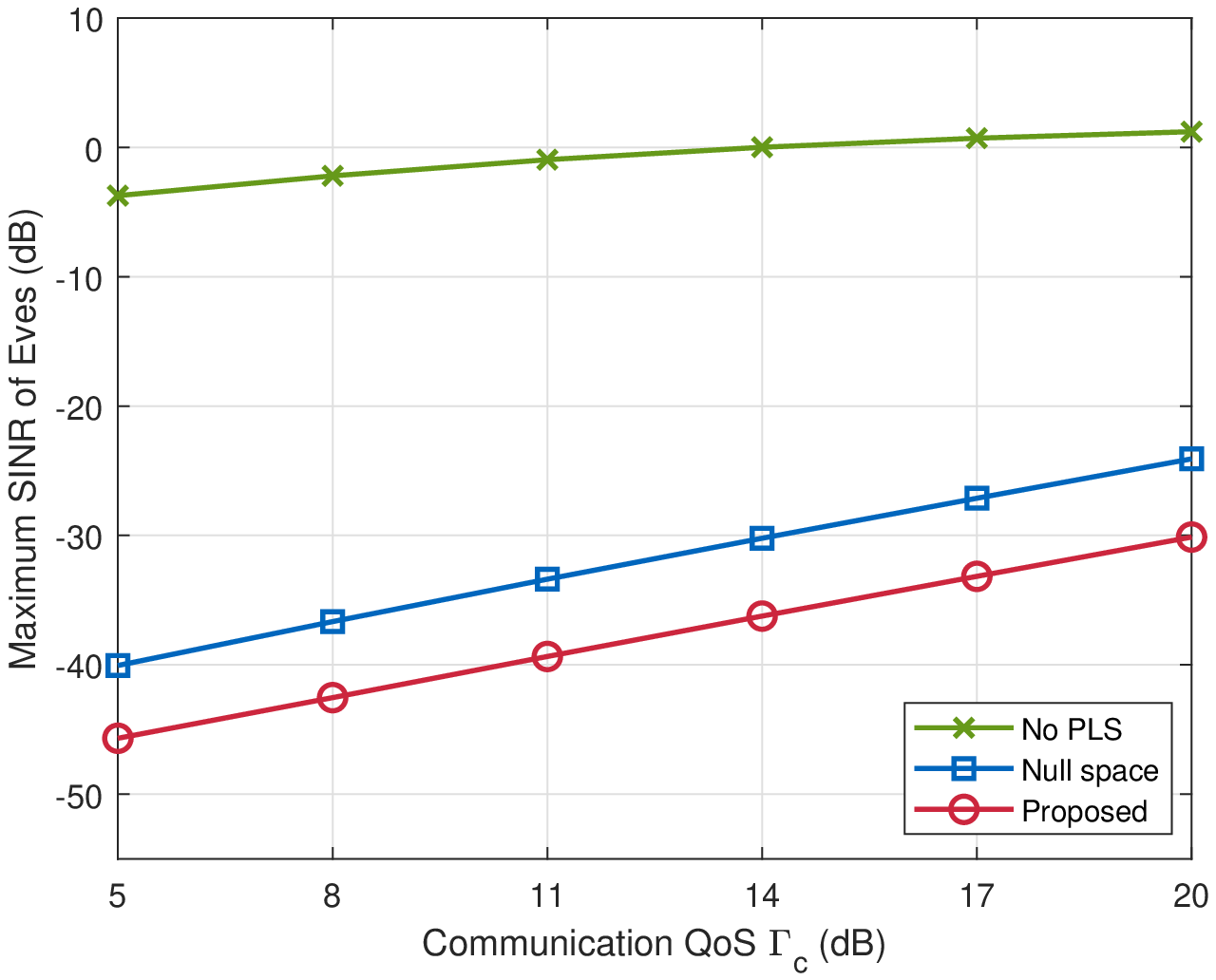}%\vspace{-0.1 cm}
\caption{SINR of Eves versus the communication QoS requirement $\Gamma_\mathrm{c}$ ($M$ = 16, $\Gamma_\mathrm{r}$ = 10dB, $P_\mathrm{r}$ = 1000W).}\label{fig:part2_gamma} %\vspace{-0.3 cm}
\end{figure}

\begin{figure}[t]
\centering
\includegraphics[width = 3.6 in]{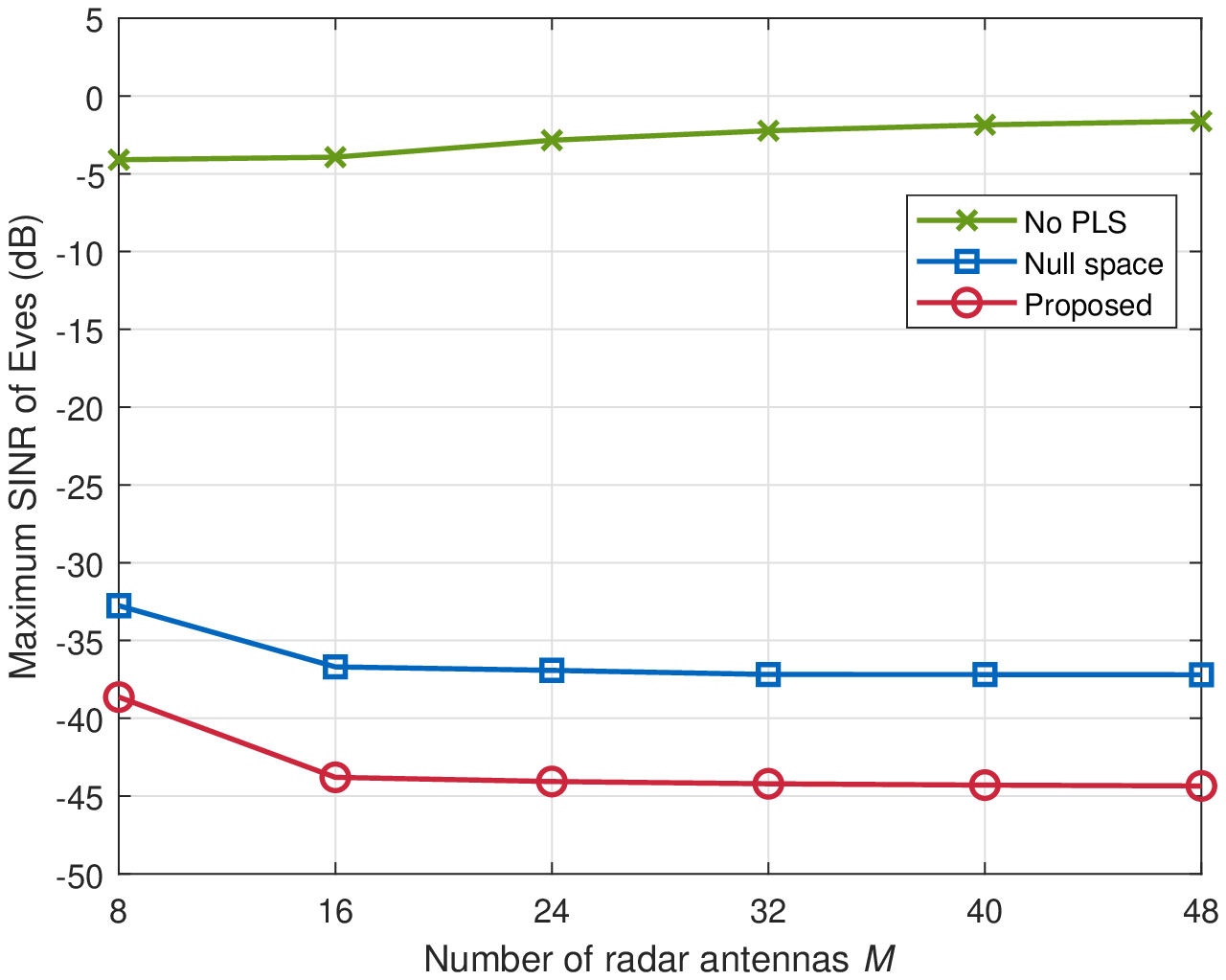}%\vspace{-0.1 cm}
\caption{SINR of Eves versus the number of radar antennas $M$ ($\Gamma_\mathrm{c}$ = 5dB, $\Gamma_\mathrm{r}$ = 10dB, $P_\mathrm{r}$ = 500W).}\label{fig:part2_M} %\vspace{-0.3 cm}
\end{figure}

\begin{figure}[t]
\centering
\includegraphics[width = 3.6 in]{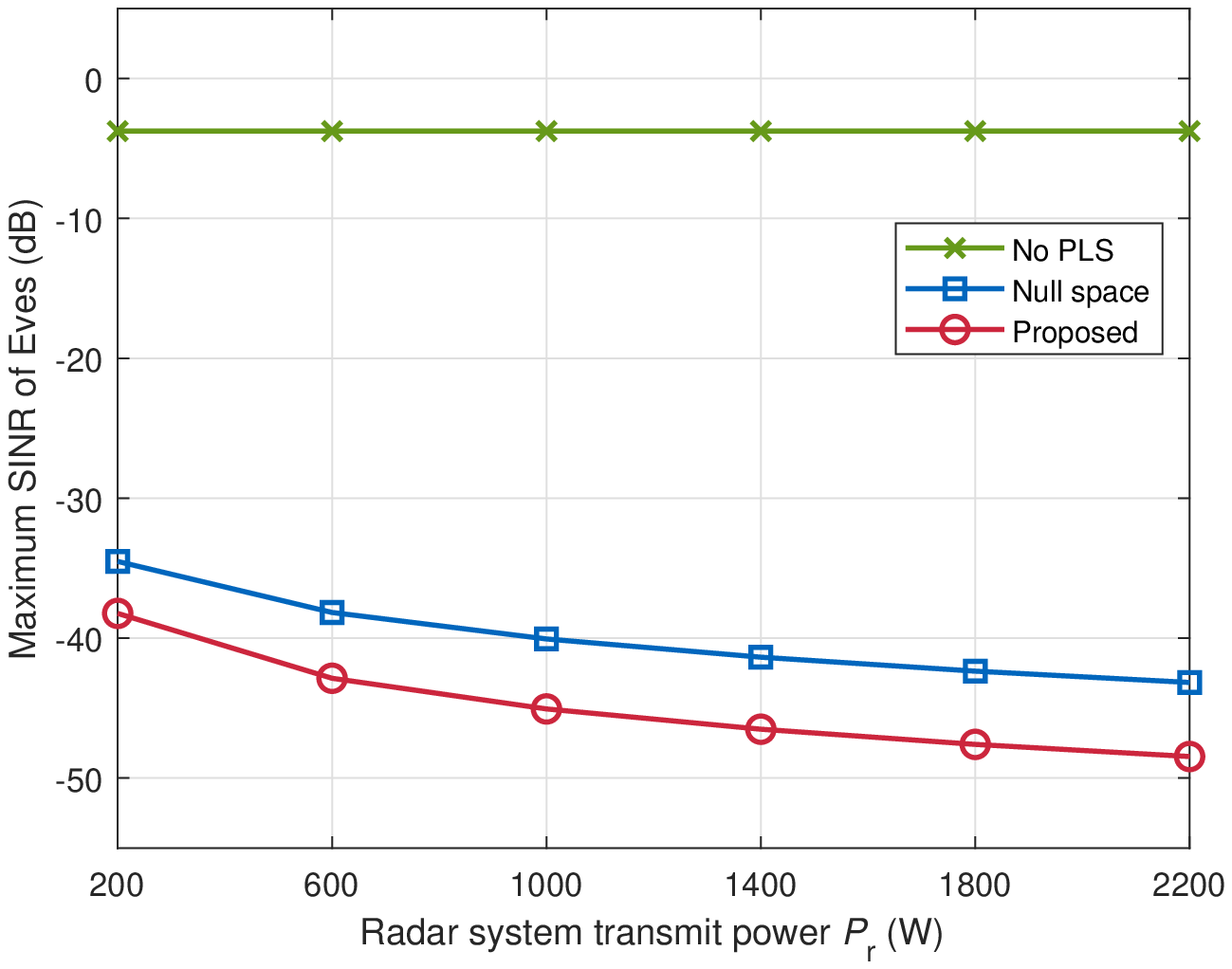}%\vspace{-0.1 cm}
\caption{SINR of Eves versus the radar system transmit power $P_\mathrm{r}$ ($M$ = 16, $\Gamma_\mathrm{c}$ = 5dB, $\Gamma_\mathrm{r}$ = 10dB).}\label{fig:part2_Pr} %\vspace{-0.3 cm}
\end{figure}

Next, we present the security performance of the proposed algorithm in Figs. \ref{fig:part2_gamma}-\ref{fig:part2_Pr}.
We also include a scheme that does not consider the physical layer security.
In other words, it only uses the least power to satisfy the communication QoS requirements of the legitimate users and the radar output SINR constraint (denoted as ``No PLS''), and the scheme that utilizes the tremendous residual power to generate AN in the null space of Bobs' channels (denoted as ``Null space'').
The security of the communication system is evaluated in terms of the maximum eavesdropping SINR of the potential eavesdroppers on the $K$ communication users.

%% \gamma

%% M

%% P_r

In Fig. \ref{fig:part2_gamma}, we plot the maximum SINR of Eves versus the communication QoS requirement $\Gamma_\mathrm{c}$.
Not surprisingly, the maximum SINR of Eves increases with increasing communication QoS requirements, because the stricter communication QoS requirements require more power for the beamforming and leave less power to generate AN, which brings higher risks of being eavesdropped.
In addition, we observe a considerable performance gap between the ``No PLS'' scheme and the schemes with PLS design (``Null space'' and ``Proposed''), since the generated AN greatly confuses the eavesdropper.
Moreover, it is seen that the proposed algorithm achieves notably better security performance than the ``Null space'' scheme since more DoFs are exploited in designing AN.
These findings verify the advantages of the proposed AN-aided transmit beamforming design in preventing potential eavesdropping without the knowledge of Eves' CSI.

Fig. \ref{fig:part2_M} shows the maximum SINR of Eves versus the number of radar antennas $M$.
Since more antennas can exploit additional spatial DoFs, the transmit power required to achieve the pre-set radar SINR becomes less.
Therefore, for the ``No PLS'' scheme, the jamming/interference from the radar to the eavesdropper becomes weak and the security performance is degraded.
On the other hand, for the ``Null space'' scheme and the proposed scheme, more power is available for generating AN, which can significantly improve the security performance.
It is noticed that, compared with Fig. 4, Eves' SINR will not visibly decrease when the number of radar antennas is greater than 16.
This is because the spatial DoFs of a radar system with 16 antennas are sufficient for providing enough signal processing ability.
More radar antennas would not significantly reduce the transmit power for achieving the required radar SINR. Consequently, the available power for generating AN will be relatively small with more radar antenna than 16, and the Eves' SINR will be maintained at almost the same extremely low level, e.g. close to -44dB. These results can still verify that the proposed AN-aided secure beamforming design can provide superior security performance even when the MIMO radar has a moderate number of antennas.

Fig. \ref{fig:part2_Pr} presents the maximum SINR of Eve versus the radar system power budget $P_\mathrm{r}$.
The same relationship between different design schemes can be observed in Figs. \ref{fig:part2_gamma} and \ref{fig:part2_M}.
In addition, since more additional power can be utilized for generating AN, the eavesdropping SINR decreases with the increase of $P_\mathrm{r}$ for the ``Null space'' and the proposed schemes.

%\vspace{-0.1cm}
\section{Conclusions}

This paper investigated joint secure transmit beamforming designs for ISAC systems.
When the eavesdroppers' channels are available, the maximum eavesdropping SINR was minimized under the communication QoS constraints, the radar detection performance constraint, and the communication and radar power constraints.
An efficient BCD-FP-SDR-based algorithm was proposed to solve the non-convex optimization problem.
When the eavesdroppers' channels are unavailable, a joint AN-aided transmit beamforming design was developed to disrupt the eavesdroppers' reception while guaranteeing the legitimate communication SINR and the radar output SINR requirements by utilizing the available power of radar and communication systems to generate as much AN as possible.
We proposed a double-loop BCD-SDR-based algorithm to solve the resulting non-convex optimization problem.
Simulation results illustrated the advantages of ISAC systems on secure transmissions and the effectiveness of the proposed algorithms.

\begin{appendices}
\section{}

Let $\widetilde{\mathbf{W}}_{k}$, $\forall k$, be an arbitrary global optimal solution to problem \eqref{eq:SDR}.
Then, we construct a new solution $\widehat{\mathbf{W}}_{k}$, $\forall k$, based on $\widetilde{\mathbf{W}}_{k}$, $\forall k$, as
\begin{equation}
\label{eq:construct vk}
\widehat{\mathbf{W}}_{k}=\widetilde{\mathbf{W}}_{k},
~\widehat{\mathbf{w}}_{k}=(\mathbf{h}_k^H\widetilde{\mathbf{W}}_{k}\mathbf{h}_k)^{-1/2}\widetilde{\mathbf{W}}_{k}\mathbf{h}_k,
~\widehat{\mathbf{W}}_{k}=\widehat{\mathbf{w}}_{k}\widehat{\mathbf{w}}_{k}^H.
\end{equation}
Clearly, $\widehat{\mathbf{W}}_{k}$, $\forall k$, is rank-one and positive semidefinite which means that constraints (\ref{eq:SDR}g) and (\ref{eq:SDR}h) hold.
Next, we show that $\widehat{\mathbf{W}}_{k}$ is also a global optimal solution to problem \eqref{eq:SDR}.
Based on \eqref{eq:construct vk}, we have the following equality
\begin{equation}
\label{eq:wk equality}
\sum_{k=1}^{K}\widetilde{\mathbf{W}}_{k}=\sum_{k=1}^{K}\widehat{\mathbf{W}}_{k},
\end{equation}
which implies that the constraints (\ref{eq:SDR}d) and (\ref{eq:SDR}e) hold.
Moreover, the constraint (\ref{eq:SDR}f) and the objective function hold since they do not contain the variables $\widetilde{\mathbf{W}}_{k}$, $\forall k$.

In addition, since
\begin{equation}
\label{eq:hkWk}
\mathbf{h}_k^H\widehat{\mathbf{W}}_{k}\mathbf{h}_k
=\mathbf{h}_k^H\widehat{\mathbf{w}}_{k}\widehat{\mathbf{w}}_{k}^H\mathbf{h}_k
=\mathbf{h}_k^H\widetilde{\mathbf{W}}_{k}\mathbf{h}_k,
\end{equation}
the SINR constraint in (\ref{eq:SDR}c) can be re-written as
\begin{equation}
\begin{aligned}
&\frac{(\Gamma_k+1)}{\Gamma_k}\mathbf{h}_k^H\widehat{\mathbf{W}}_{k}\mathbf{h}_k =
\frac{(\Gamma_k+1)}{\Gamma_k}\mathbf{h}_k^H\widetilde{\mathbf{W}}_{k}\mathbf{h}_k\\
&\geq \sum_{j=1}^{K}\mathbf{h}_k^H\widetilde{\mathbf{W}}_{j}\mathbf{h}_k + \mathbf{g}_k^H \mathbf{R}_\mathrm{F}\mathbf{g}_k + \sigma_{k}^{2} \\
&= \sum_{j=1}^{K}\mathbf{h}_k^H\widehat{\mathbf{W}}_{j}\mathbf{h}_k + \mathbf{g}_k^H \mathbf{R}_\mathrm{F}\mathbf{g}_k + \sigma_{k}^{2},
\end{aligned}
\end{equation}
namely (\ref{eq:SDR}c) holds for $\widehat{\mathbf{W}}_{k}$, $\forall k$.
Similarly, the SINR constraint (\ref{eq:SDR}b) also holds for $\widehat{\mathbf{W}}_{k}$, $\forall k$.
Therefore, it is proven that the constructed $\widehat{\mathbf{W}}_{k}$, $\forall k$ in \eqref{eq:construct vk} is a feasible and global optimal rank-1 solution to the problem (\ref{eq:SDR}).

For problem \eqref{eq:calculate Wk and Rz}, we also let $\widetilde{\mathbf{W}}_{k}$, $\forall k$, be an arbitrary global optimal solution to problem \eqref{eq:calculate Wk and Rz}.
In the same way as \eqref{eq:construct vk}, we construct a new solution $\widehat{\mathbf{W}}_{k}$, $\forall k$, which implies that (\ref{eq:calculate Wk and Rz}e) holds.
Similar to the previous proof, constraints (\ref{eq:calculate Wk and Rz}a)-(\ref{eq:calculate Wk and Rz}d) hold.
Thus, the constructed $\widehat{\mathbf{W}}_{k}$, $\forall k$, in \eqref{eq:construct vk} being a feasible and global optimal solution to problem \eqref{eq:calculate Wk and Rz} is verified.
\end{appendices}

%\vspace{-0.05cm}
%参考文献

\end{document}